\documentclass{article}
\usepackage{fullpage}

\newcommand{\onlyarxiv}[1]{{#1}}

\usepackage{wrapfig}

\makeatletter
\newcommand\gobblepars{%
    \@ifnextchar\par%
        {\expandafter\gobblepars\@gobble}%
        {}}
\makeatother
\renewcommand{\paragraph}[1]{\smallskip\noindent\textbf{#1}.\ \ \gobblepars}

\newcommand{\tightlinespace}{-1.3ex}

\newtheorem{preexperiment}{Experiment}

\usepackage{balance}
\usepackage{amsmath}
\usepackage{amsfonts}
\usepackage{amssymb}
\usepackage{ellipsis} %
\usepackage[hyphens]{url} %
\usepackage{verbatim} %
\usepackage{epsfig,color} %
\usepackage{stmaryrd} %
\usepackage{multirow} %

\newcommand{\ftnote}{\footnote}

\newcommand*{\set}[2]{\{\,{#1}~|~{#2}\,\}}

\newcommand*{\given}{\mathop{{|}}}

\newcommand{\mc}{\mathcal}
\newcommand{\bigtimes}{\times}

\usepackage{booktabs} %
\newenvironment{tab}[1]
{%
\let\oldarraystretch=\arraystretch
\renewcommand{\arraystretch}{1.2} %
\mbox{}\hfill
\begin{tabular}{@{}#1@{}}
\toprule
}
{\bottomrule
\end{tabular}
\mbox{}\hfill
\renewcommand{\arraystretch}{\oldarraystretch}
}

\newcommand{\midruleheaderbottom}{\hline}

\newenvironment{tablewide}{\begin{table}\footnotesize}{\end{table}}

\usepackage{array}
\newcolumntype{L}[1]{>{\raggedright\let\newline\\\arraybackslash\hspace{0pt}}m{#1}}
\newcolumntype{C}[1]{>{\centering\let\newline\\\arraybackslash\hspace{0pt}}m{#1}}
\newcolumntype{R}[1]{>{\raggedleft\let\newline\\\arraybackslash\hspace{0pt}}m{#1}}

\newcommand{\email}{\url}

\author{%
Amit Datta\\
Carnegie Mellon University\\
\email{amitdatta@cmu.edu}
\and
Michael Carl Tschantz\\
International Computer Science Institute\\
\email{mct@icsi.berkeley.edu}
\and
Anupam Datta\\
Carnegie Mellon University\\
\email{danupam@cmu.edu}
}

\title{Automated Experiments on Ad Privacy Settings\\ A Tale of Opacity, Choice, and Discrimination\thanks{This report is an extended version of an article appearing in the Proceedings on Privacy Enhancing Technologies~\cite{datta15pets}.  Both of those works build upon an earlier version of this report~\cite{datta14arxiv}.}}

\begin{document}

\maketitle

\begin{abstract}
To partly address people's concerns over web tracking, Google has created the Ad Settings webpage to provide information about and some choice over the profiles Google creates on users.  We present AdFisher, an automated tool that explores how user behaviors, Google's ads, and Ad Settings interact. AdFisher can run browser-based experiments and analyze data using machine learning and significance tests. 
Our tool uses a rigorous experimental design and statistical analysis to ensure the statistical soundness of our results.  
We use AdFisher to find that the Ad Settings was opaque about some features of a user's profile, that it does provide some choice on ads, and that these choices can lead to seemingly discriminatory ads.  In particular, we found that visiting webpages associated with substance abuse changed the ads shown but not the settings page.  We also found that setting the gender to female resulted in getting fewer instances of an ad related to high paying jobs than setting it to male.  We cannot determine who caused these findings due to our limited visibility into the ad ecosystem, which includes Google, advertisers, websites, and users. Nevertheless, these results can form the starting point for deeper investigations by either the companies themselves or by regulatory bodies.
\end{abstract}

\section{Introduction}

\paragraph{Problem and Overview}
With the advancement of tracking technologies and the growth of online data aggregators, data collection on the Internet has become a serious privacy concern. Colossal amounts of collected data are used, sold, and resold for serving targeted content, notably advertisements, on websites
(e.g.,~\cite{mayer12sp}).  Many websites providing content, such as news, outsource their advertising operations to large third-party ad networks, such as Google's DoubleClick.  These networks embed tracking code into webpages across many sites providing the network with a more global view of each user's behaviors. 

\begin{wrapfigure}{r}{10.1cm}
\centering
\includegraphics[width=10cm]{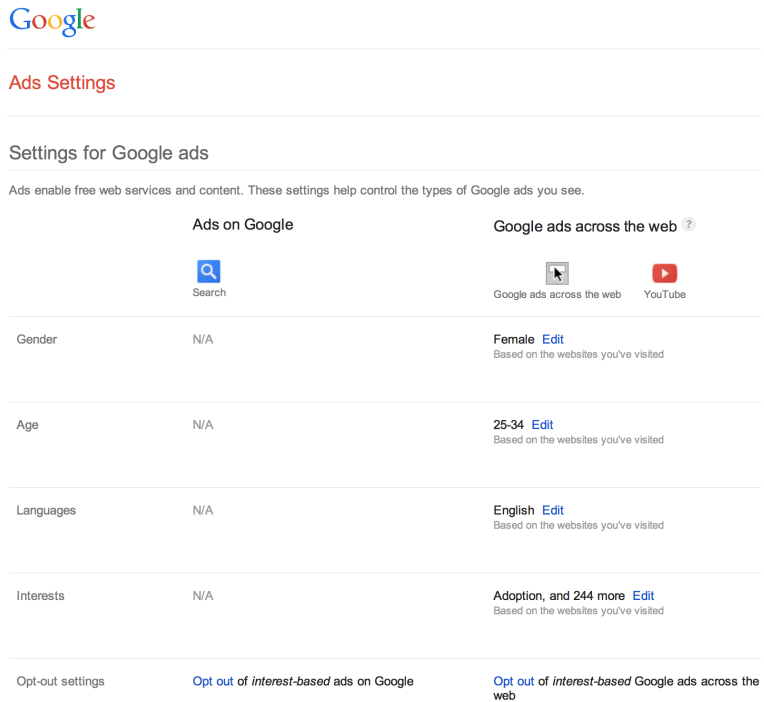}
\caption{Screenshot of Google's Ad Settings webpage}
\label{fig:settings}
\end{wrapfigure}

People are concerned about behavioral marketing on the web~(e.g.,~\cite{ur12soups}). %
To increase transparency and control, %
Google provides Ad Settings, which is ``a Google tool that helps you control the ads you see on Google services and on websites that partner with Google''~\cite{google-ad-settings-help}. It displays inferences Google has made about a user's demographics and interests based on his browsing behavior.
Users can view and edit these settings at\\
\centerline{\url{http://www.google.com/settings/ads}} %
Figure~\ref{fig:settings} provides a screenshot.
Yahoo~\cite{yahoo-help}
and Microsoft~\cite{microsoft-choice} also offer personalized ad settings.

However, they provide little information about how these pages operate, leaving open the question of
how completely these settings describe the profile they have about a user. In this study, we explore how a user's behaviors, either directly with the settings or with content providers, alter the ads and settings shown to the user and whether these changes are in harmony. 
In particular, we study the degree to which the settings provides transparency and choice as well as checking for the presence of discrimination.
Transparency is important for people to understand how the use of data about them affects the ads they see.
Choice allows users to control how this data gets used, enabling them to protect the information they find sensitive.
Discrimination is an increasing concern about machine learning systems and one reason people like to keep information private~\cite{bigdata14whitehouse, zemel13icml}.

To conduct these studies, we developed AdFisher, a tool for automating randomized, controlled experiments for studying online tracking.
Our tool offers a combination of automation, statistical rigor, scalability, and explanation for determining the use of information by web advertising algorithms and by personalized ad settings, such as Google Ad Settings.
The tool can simulate having a particular interest or attribute by visiting webpages associated with that interest or by altering the ad settings provided by Google.  It collects ads served by Google and also the settings that Google provides to the simulated users. It automatically analyzes the data to determine whether statistically significant differences between groups of agents exist.  AdFisher uses machine learning to automatically detect differences and then executes a test of significance specialized for the difference it found.

Someone using AdFisher to study behavioral targeting only has to provide the behaviors the two groups are to perform (e.g., visiting websites) and the measurements (e.g., which ads) to collect afterwards. 
AdFisher can easily run multiple experiments exploring the causal connections between users' browsing activities, and the ads and settings that Google shows.

The advertising ecosystem is a vast, distributed, and decentralized system with several players including the users consuming content, the advertisers, the publishers of web content, and ad networks.  With the exception of the user, we treat the entire ecosystem as a blackbox.  We measure simulated users' interactions with this blackbox including page views, ads, and ad settings.  
Without knowledge of the internal workings of the ecosystem, we cannot assign responsibility for our findings to any single player within it nor rule out that they are unintended consequences of interactions between players.
However, our results show the presence of concerning effects illustrating the existence of issues that could be investigated more deeply by either the players themselves or by regulatory bodies with the power to see the internal dynamics of the ecosystem.

\paragraph{Motivating Experiments}
In one experiment, we explored whether visiting websites related to substance abuse has an impact on Google's ads or settings.  We created an experimental group and a control group of agents.  The browser agents in the experimental group visited  websites on substance abuse while the agents in the control group simply waited.  Then, both groups of agents collected ads served by Google on a news website. %

Having run the experiment and collected the data, we had to determine whether any difference existed in the outputs shown to the agents.  One way would be to intuit what the difference could be (e.g. more ads containing the word ``alcohol'') and test for that difference.  %
However, developing this intuition %
can take considerable effort. Moreover, it does not help find unexpected differences. Thus, we instead used
machine learning to automatically find differentiating patterns in %
the data. Specifically, AdFisher finds a classifier that can predict which group an agent belonged to, from the ads shown to an agent. %
The classifier is trained on a subset of the data. A separate test subset is used to determine whether the classifier found a statistically significant difference between the ads shown to each group of agents.
In this experiment, AdFisher found a classifier that could distinguish between the two groups of agents by using the fact that only the agents that visited the substance abuse websites received ads for Watershed Rehab.

We also measured the settings that Google provided to each agent on its Ad Settings page after the experimental group of agents visited the webpages associated with substance abuse.  We found no differences (significant or otherwise) between the pages for the agents.  Thus, information about visits to these websites is indeed being used to serve ads, but the Ad Settings page does not reflect this use in this case. %
Rather than providing transparency, in this instance, the ad settings were \emph{opaque} as to the impact of this factor.

In another experiment, we examined whether the settings provide \emph{choice} to  users.  We found that removing interests from the Google Ad Settings page changes the ads that a user sees.  In particular, we had both groups of agents visit a site related to online dating.  Then, only one of the groups removed the interest related to online dating.  Thereafter, the top ads shown to the group that kept the interest were related to dating but not the top ads shown to the other group. Thus, the ad settings do offer the users a degree of choice over the ads they see.

We also found evidence suggestive of \emph{discrimination} from another experiment.
We set the agents' gender to female or male on Google's Ad Settings page.  We then had both the female and male groups of agents visit webpages associated with employment.  %
We established that Google used this gender information to select ads, as one might expect. The interesting result was how the ads differed between the groups: during this experiment, Google showed the simulated males %
ads from a certain career coaching agency that promised large salaries more frequently than the simulated females, a finding suggestive of discrimination. 
Ours is the first study that provides statistically significant evidence of an instance of discrimination in online advertising when demographic information is supplied via a transparency-control mechanism (i.e., the Ad Settings page).

While neither of our findings of opacity or discrimination are clear violations of Google's privacy policy~\cite{google-privacy} and we do not claim these findings to generalize or imply widespread issues, we find them concerning and warranting further investigation by those with visibility into the ad ecosystem.
Furthermore, while our finding of discrimination in the non-normative sense of the word is on firm statistical footing, we acknowledge that people may disagree about whether we found discrimination in the normative sense of the word.  We defer discussion of whether our findings suggest unjust discrimination until Section~\ref{sec:disc}.

\paragraph{Contributions}
In addition to the experimental findings highlighted above, we provide AdFisher, a tool for \emph{automating} such experiments.  AdFisher is structured as a Python API providing functions for setting up, running, and analyzing experiments.
We use Selenium to drive Firefox browsers and the scikit-learn library~\cite{scikit-learn} for implementations of classification algorithms. 
We use the SciPy library~\cite{scipy} for implementing the statistical analyses of the core methodology.

AdFisher offers \emph{rigor} by performing a carefully designed experiment. The statistical analyses techniques applied do not make questionable assumptions about the collected data. We base our design and analysis on a prior proposal that makes no assumptions about the data being independent or identically distributed~\cite{tschantz14arxiv}. Since advertisers update their behavior continuously in response to unobserved inputs (such as ad auctions) and the experimenters' own actions, such assumptions may not always hold. %
Indeed, in practice, the distribution of ads changes over time and simulated users, or \emph{agents}, interfere with one another~\cite{tschantz14arxiv}.

Our automation, experimental design, and statistical analyses allow us to \emph{scale} to handling large numbers of agents for finding subtle differences.  In particular, we modify the prior analysis of Tschantz et al.~\cite{tschantz14arxiv} to allow for experiments running over long periods of time.  We do so by using \emph{blocking} (e.g.,~\cite{good05book}), a nested statistical analysis not previously applied to understanding web advertising.  The blocking analysis ensures that agents are only compared to the agents that start out like it and then aggregates together the comparisons across blocks of agents.  Thus, AdFisher may run agents in batches spread out over time while only comparing those agents running simultaneously to one another.

AdFisher also provides \emph{explanations} as to how Google alters its behaviors in response to different user actions. 
It uses the trained classifier model to find which features were most useful for the classifier to make its predictions. It provides the top features from each group to provide the experimenter/analyst with a qualitative understanding of how the ads differed between the groups. %
To maintain statistical rigor, we carefully circumscribe our claims. We only claim statistical soundness of our results: if our techniques detect an effect of the browsing activities on the ads, then there is indeed one with high likelihood (made quantitative by a p-value). We do not claim that we will always find a difference if one exists, nor that the differences we find are typical of those experienced by users. Furthermore, while we can characterize the differences, we cannot assign blame for them since either Google or the advertisers working with Google could be responsible.

\paragraph{Contents}
After covering prior work next, we present, in Section~\ref{sec:prop}, privacy properties that our tool AdFisher can check: nondiscrimination, transparency, and choice.
Section~\ref{sec:meth} explains the methodology we use to ensure sound conclusions from using AdFisher.  Section~\ref{sec:tool} presents the design of AdFisher.  Section~\ref{sec:expr} discusses our use of AdFisher to study Google's ads and settings.
We end with conclusions and future work.

Raw data and additional details about AdFisher and our experiments can be found at\\
\centerline{\url{http://www.cs.cmu.edu/~mtschant/ife/}}
AdFisher is freely available at \\
\centerline{\url{https://github.com/tadatitam/info-flow-experiments/}}

\section{Prior Work}

We are not the first to study how Google uses information.
The work with the closest subject of study to ours is by Wills and Tatar~\cite{wills12wpes}.  They studied both the ads shown by Google and the behavior of Google's Ad Settings (then called the ``Ad Preferences'').  Like us, they find the presence of opacity: various interests impacted the ads and settings shown to the user and that ads could change without a corresponding change in Ad Settings.  
Unlike our study, theirs was mostly manual, small scale, lacked any statistical analysis, and did not follow a rigorous experimental design.  
Furthermore, we additionally study choice and discrimination.

Other related works differ from us in both goals and methods.  They all focus on how visiting webpages change the ads seen.  While we examine such changes in our work, we do so as part of a larger analysis of the interactions between ads and personalized ad settings, a topic they do not study.

Barford et al.\ come the closest in that their recent study looked at both ads and ad settings~\cite{barford14www}.
They do so in their study of the ``adscape'', an attempt to understand each ad on the Internet.
They study each ad individually and cast a wide net to analyze many ads from many websites while simulating many different interests.
They only examine the ad settings to determine whether they successfully induced an interest.
We rigorously study how the settings affects the ads shown (choice) and how behaviors can affect ads without affecting the settings (transparency).
Furthermore, we use focused collections of data and an analysis that considers all ads collectively to find subtle causal effects within Google's advertising ecosystem.
We also use a %
randomized experimental design and analysis to ensure that our results imply causation.

The usage study closest to ours in statistical methodology is that of Tschantz et al.~\cite{tschantz14arxiv}. 
They developed a rigorous methodology for determining whether a system like Google uses information.
Due to limitations of their methodology, they only ran small-scale studies.
While they observed that browsing behaviors could affect Ad Settings, they did not study how this related to the ads received.
Furthermore, while we build upon their methodology, we automate the selection of an appropriate test statistic by using machine learning whereas they manually selected test statistics.

The usage study closest to ours in terms of implementation is that of Liu et al.\ in that they also use machine learning~\cite{liu13hotnets}.
Their goal is to determine whether an ad was selected due to the content of a page, by using behavioral profiling, or from a previous webpage visit.
Thus, rather than using machine learning to select a statistical test for finding causal relations, they do so to detect whether an ad on a webpage matches the content on the page to make a case for the first possibility.  Thus, they have a separate classifier for each interest a webpage might cover.  Rather than perform a statistical analysis to determine whether treatment groups have a statistically significant difference, they use their classifiers to judge the ratio of ads on a page unrelated to the page's content, which they presume indicates that the ads were the result of behavioral targeting.

L\'ecuyer et al.\ present XRay, a tool that looks for correlations between the data that web services have about users and the ads shown to users~\cite{lecuyer14usenix}.  
While their tool may check many changes to a type of input to determine whether any of them has a correlation with the frequency of a single ad, it does not check for causation, as ours does.

Englehardt et al.\ study filter bubbles with an analysis that assumes independence between observations~\cite{englehardt14man}, an assumption we are uncomfortable making.  (See Section~\ref{sec:pitfalls}.)

Guha et al.\ compare ads seen by three agents to see whether Google treats differently the one that behaves differently from the other two~\cite{guha10imc}.  We adopt their suggestion of focusing on the title and URL displayed on ads when comparing ads to avoid noise from other less stable parts of the ad.  Our work differs by studying the ad settings in addition to the ads and by using larger numbers of agents.  Furthermore, we use rigorous statistical analyses. %
Balebako et al.\ run similar experiments 
to study the effectiveness of privacy tools~\cite{balebako12w2sp}.

Sweeney ran an experiment to determine that searching for names associated with African-Americans produced more search ads suggestive of an arrest record than names associated with European-Americans~\cite{sweeney13cacm}.  
Her study required considerable insight to determine that suggestions of an arrest was a key difference.
AdFisher can automate not just the collection of the ads, but also the identification of such key differences by using its machine learning capabilities.  Indeed, it found on its own that simulated males were more often shown ads encouraging the user to seek coaching for high paying jobs than simulated females.

\section{Privacy Properties}
\label{sec:prop}

Motivating our methodology for finding causal relationships, we present some properties of ad networks that we can check with such a methodology in place.
As a fundamental limitation of science, we can only prove the existence of a causal effect; we cannot prove that one does not exist (see Section~\ref{sec:scope}).  
Thus, experiments can only demonstrate violations of nondiscrimination and transparency, which require effects.
On the other hand, we can experimentally demonstrate that effectful choice and ad choice are complied with in the cases that we test since compliance follows from the existence of an effect.
Table~\ref{tbl:properties} summarizes these properties.

\begin{tablewide}
\newcommand*{\extrarowgap}{2.4ex}
\begin{tab}{lL{5cm}L{5cm}l}
Property Name & Requirement & Causal Test & Finding\\
\midrule
Nondiscrimination & Users differing only on protected attributes are treated similarly & Find that presence of protected attribute causes a change in ads & Violation\\[\extrarowgap]
Transparency & User can view all data about him used for ad selection & Find  attribute that causes a change in ads, not in settings & Violation\\[\extrarowgap]
Effectful choice & Changing a setting has an effect on ads & Find that changing a setting causes a change in ads & Compliance\\[\extrarowgap]
Ad choice & Removing an interest decreases the number ads related to that interest & Find setting causes a decease in relevant ads & Compliance\\
\end{tab}
\caption{Privacy Properties Tested on Google's Ad Settings}
\label{tbl:properties}
\end{tablewide}

\subsection{Discrimination}

At its core, \emph{discrimination} between two classes of individuals (e.g., one race vs.\ another) occurs when the attribute distinguishing those two classes causes a change in behavior toward those two classes.
In our case, discrimination occurs when membership in a class causes a change in ads.
Such discrimination is not always bad (e.g., many would be comfortable with men and women receiving different clothing ads).  We limit our discussion of whether the discrimination we found is unjust to the discussion section (\S\ref{sec:disc}) and do not claim to have a scientific method of determining the morality of discrimination.

Determining whether class membership causes a change in ads is difficult since many factors not under the experimenter's control or even observable to the experimenter may also cause changes.  
Our experimental methodology determines when membership in certain classes causes significant changes in ads by comparing many instances of each class.

We are limited in the classes we can consider since we cannot create actual people that vary by the traditional subjects of discrimination, such as race or gender. 
Instead, we look at classes that function as surrogates for those classes of interest.  
For example, rather than directly looking at how gender affects people's ads, we instead look at how altering a gender setting affects ads or at how visiting websites associated with each gender affects ads.

\subsection{Transparency}

Transparency tools like Google Ad Settings provide online consumers with some understanding of the information that ad networks collect and use about them. By displaying to users what the ad network may have learned about the interests and demographics of a user, such tools attempt to make targeting mechanisms more transparent. %

However the technique for studying transparency is not clear. One cannot expect an ad network to be \emph{completely transparent} to a user. This would involve the tool displaying all other users' interests as well. A more reasonable expectation is for the ad network to display any inferred interests about that user. So, if an ad network has inferred some interest about a user and is serving ads relevant to that interest, then  that interest should be displayed on the transparency tool. 
However, even this notion of transparency cannot be checked precisely as the ad network may serve ads about some other interest correlated with the original inferred interest, but not display the correlated interest on the transparency tool. %

Thus, we only study the extreme case of the lack of transparency --- \emph{opacity}, and leave complex notions of transparency open for future research. We say that a transparency tool has opacity if some browsing activity results in 
a significant effect on the ads served, but has no effect on the ad settings. If there is a difference in the ads, we can argue that prior browsing activities must have been tracked and used by the ad network to serve relevant ads. However, if this use does not show up on the transparency tool, we have found at least one example which %
demonstrates a lack of transparency. 

\subsection{Choice}

The Ad Settings page offers users the option of editing the interests and demographics inferred about them.
However, the exact nature of how these edits impact the ad network is unclear.  
We examine two notions of choice.  

A very coarse form is \emph{effectful choice}, which requires that altering the settings has some effect on the ads seen by the user.  
This shows that altering settings is not merely a ``placebo button'': it has a real effect on the network's ads. However, effectful choice does not capture whether the effect on ads is meaningful. For example, even if a user adds interests for cars and starts receiving \emph{fewer} ads for cars, effectful choice is satisfied. 
Moreover, we cannot find violations of effectful choice. 
If we find no differences in the ads, we cannot conclude that users do not have effectful choice since it could be the result of the ad repository lacking ads relevant to the interest.

Ideally, the effect on ads after altering a setting would be meaningful and related to the changed setting.
One way such an effect would be meaningful, in the case of removing an inferred interest, is a decrease in the number of ads related to the removed interest. %
We call this requirement \emph{ad choice}.
One way to judge whether an ad is relevant is to check it for keywords associated with the interest.  
If upon removing an interest, we find a statistically significant decrease in the number of ads containing some keywords, then we will conclude that the choice was respected.
In addition to testing for compliance in ad choice, we can also test for a violation by checking for a statistically significant increase in the number of related ads to find egregious violations.
By requiring the effect to have a fixed direction, we can find both compliance and violations of ad choice.

\section{Methodology}
\label{sec:meth}

The goal of our methodology is to establish that a certain type of input to a system causes an effect on a certain type of output of the system.  For example, in our experiments, we study the system of Google.  The inputs we study are visits to content providing websites and users' interactions with the Ad Settings page.  The outputs we study are the settings and ads shown to the users by Google.  However, nothing in our methodology limits ourselves to these particular topics; it is appropriate for determining I/O properties of any web system.
Here, we present an overview of our methodology; Appendix~\ref{app:stat} provides details of the statistical analysis.
\subsection{Background: Significance Testing}

To establish causation, we start with the approach of Fisher (our tool's namesake) for significance testing~\cite{fisher35doe} as specialized by Tschantz et al.\ for the setting of online systems~\cite{tschantz14arxiv}.
Significance testing examines a \emph{null hypothesis}, in our case, that the inputs do not affect the outputs. To test this hypothesis the experimenter selects two values that the inputs could take on, typically called the \emph{control} and \emph{experimental} \emph{treatments}.
The experimenter applies the treatments to \emph{experimental units}.  In our setting, the units are the browser agents, that is, simulated users. 
To avoid noise, the experimental units should initially be as close to identical as possible as far as the inputs and outputs in question are concerned.
For example, an agent created
\begin{wrapfigure}{}{8.1cm}
\centering
\includegraphics[width = 8cm]{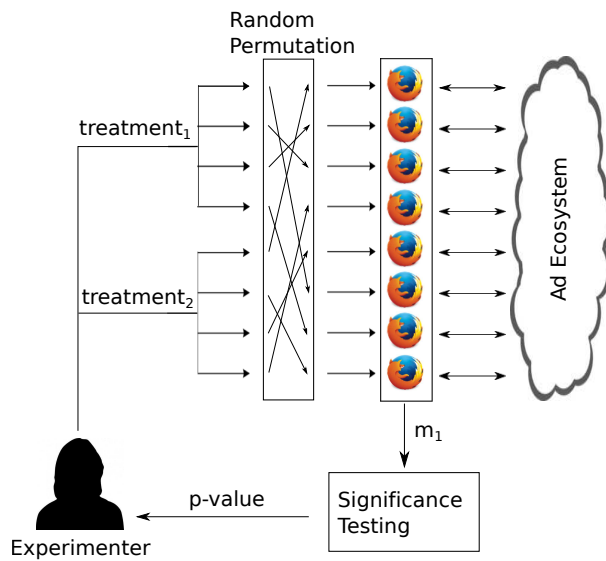}
\caption{Experimental setup to carry out significance testing on eight browser agents comparing the effects of two treatments. Each agent is randomly assigned a treatment which specifies what actions to perform on the web. After these actions are complete, they collect measurements which are used for significance testing.}
\label{fig:block}
\end{wrapfigure}
 with the Firefox browser should not be compared to one created with the Internet Explorer browser since Google can detect the browser used.

The experimenter randomly applies the experimental (control) treatment to half of the agents, which form the experimental (control) group. %
(See Figure~\ref{fig:block}.) %
Each agent carries out actions specified in the treatment applied to it. Next, the experimenter takes measurements of the outputs Google sends to the agents, such as ads.  At this point, the experiment is complete and data analysis begins.

Data analysis starts by computing a \emph{test statistic} over the measurements.  The experimenter selects a test statistic that she suspects will take on a high value when the outputs to the two groups differ. That is, the statistic is a measure of distance between the two groups.  She then uses the \emph{permutation test} to determine whether the value the test statistic actually took on is higher than what one would expect by chance unless the groups actually differ.
The permutation test randomly permutes the labels (control and experimental) associated with each observation, and recomputes a hypothetical test statistic.
Since the null hypothesis is that the inputs have no effect, %
the random assignment should have no effect on the value of the test statistic.
Thus, under the null hypothesis, it is unlikely that the actual value of the test statistic is larger than the vast majority of hypothetical values.

The \emph{p-value} of the permutation test is the proportion of the permutations where the test statistic was greater than or equal to the actual observed statistic.
If the value of the test statistic is so high that under the null hypothesis it would  take on as high of a value in less than $5\%$ of the random assignments, then we conclude that the value is \emph{statistically significant} (at the $5\%$ level) and that causation is likely.  %

\subsection{Blocking}
\label{sec:meth-block}

In practice, the above methodology can be difficult to use since creating a large number of nearly identical agents might not be possible.  In our case, we could only run ten agents in parallel given our hardware and network limitations.  Comparing agents running at different times can result in additional noise since ads served to an agent change over time.
Thus, with the above methodology, we were limited to just ten comparable units.  Since some effects that the inputs have on Google's outputs can be probabilistic and subtle, they might be missed looking at so few agents.

To avoid this limitation, we extended the above methodology to handle varying units using \emph{blocking}~\cite{good05book}.
To use blocking, we created \emph{blocks} of nearly identical agents running in parallel.  
These agents differ in terms their identifiers (e.g., process id) and location in memory.  Despite the agents running in parallel, the operating system's scheduler determines the exact order in which the agents operate.
Each block's agents were randomly partitioned into the control and experimental groups.  
This randomization ensures that the minor differences between agents noted above should have no systematic impact upon the results: these differences become noise that probably disappears as the sample size increases.
Running these blocks in a staged fashion, the experiment proceeds on block after block.  A modified permutation test now only compares the actual value of the test statistic to hypothetical values computed by reassignments of agents that respect the blocking structure.  These reassignments do not permute labels across blocks of observations.

Using blocking, we can scale to any number of agents by running as many blocks as needed.  However, the computation of the permutation test increases exponentially with the number of blocks.  Thus, rather than compute the exact p-value, we estimate it by randomly sampling the possible reassignments.  
\begin{wrapfigure}{}{8.1cm}
\centering
\includegraphics[width = 8cm]{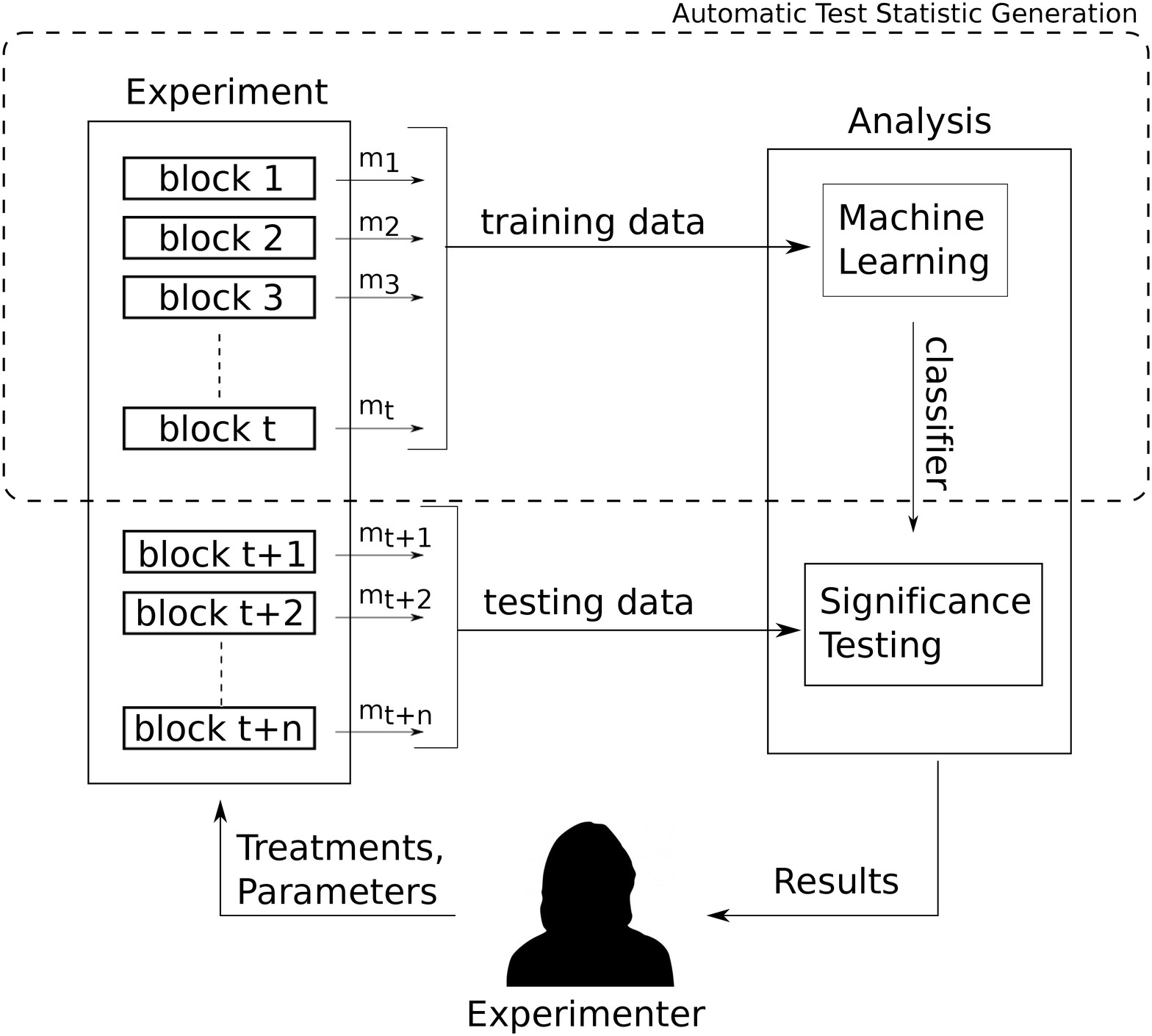}
\caption{Our experimental setup with training and testing blocks. Measurements from the training blocks are used to build a classifier. The trained classifier is used to compute the test statistic on the measurements from the testing blocks for significance testing.}\label{fig:meth2}
\onlyarxiv{\vspace{-8ex}}
\end{wrapfigure}
We can use a confidence interval to characterize the quality of the estimation~\cite{good05book}.  The p-values we report are actually the upper bounds of the $99\%$ confidence intervals of the p-values (details in Appendix~\ref{app:stat}).

\subsection{Selecting Test Statistics}

The above methodology leaves open the question of how to select the test statistic.  In some cases, the experimenter might be interested in a particular test statistic. 
For example, an experimenter testing ad choice could use a test statistic that counts the number of ads related to the removed interest.
In other cases, the experimenter might be looking for \emph{any} effect.  %
AdFisher offers the ability to automatically select a test statistic.
To do so, it partitions the collected data into training and testing subsets, and uses the training data to train a classifier. 
Figure~\ref{fig:meth2} shows an overview of AdFisher's workflow.

To select a classifier, AdFisher uses $10$-fold cross validation on the training data to select among several possible parameters. The classifier predicts which treatment an agent received, only from the ads that get served to that agent. If the classifier is able to make this prediction with high accuracy, it suggests a systematic difference between the ads served to the two groups that the classifier was able to learn. %
If no difference exists, then we would expect the number to be near the guessing rate of $50\%$. AdFisher uses the accuracy of this classifier as its test statistic. 

To avoid the possibility of seeing a high accuracy due to overfitting, AdFisher evaluates the accuracy of the classifier on a testing data set that is disjoint from the training data set.  That is, in the language of statistics, we form our hypothesis about the test statistic being able to distinguish the groups before seeing the data on which we test it to ensure that it has predictive power.  AdFisher uses the permutation test to determine whether the degree to which the classifier's accuracy on the test data surpasses the guessing rate is statistically significant.  That is, it calculates the p-value that measures the probability of seeing the observed accuracy given that the classifier is just guessing.  If the p-value is below $0.05$, we conclude that it is unlikely that classifier is guessing and that it must be making use of some difference between the ads shown to the two groups.

\subsection{Avoiding Pitfalls}
\label{sec:pitfalls}
The above methodology avoids some pitfalls.
Most fundamentally, we use a statistical analysis whose assumptions match those of our experimental design.
Assumptions required by many statistical analyses appear unjustifiable in our setting.  
For example, many analyses assume that the agents do not interact or that the ads are independent and identically distributed (e.g.,~\cite{barford14www,englehardt14man}).  
Given that all agents receive ads from the same pool of possible ads governed by the same advertisers' budgets, these assumptions appear unlikely to hold.
Indeed, empirical evidence suggests that it does not~\cite{tschantz14arxiv}.
The permutation test, which does not require this assumption, allows us to ensure  statistical soundness of our analysis without making these assumptions~\cite{greenland86epidemiology}.

Our use of randomization implies that many factors that could be confounding factors in an unrandomized design become noise in our design (e.g.,~\cite{good05book}).  While such noise may require us to use a large sample size to find an effect, it does not affect the soundness of our analysis.

Our use of two data sets, one for training the classifier to select the test statistic and one for hypothesis testing ensures that we do not engage in overfitting, data dredging, or multiple hypothesis testing (e.g.,~\cite{mitchell97book}).  All these problems result from looking for so many possible patterns that one is found by chance.  While we look for many patterns in the training data, we only check for one in the testing data.

Relatedly, by reporting a p-value, we provide a quantitative measure of the confidence we have that the observed effect is genuine and not just by chance~\cite{jensen92phd}.
Reporting simply the classifier accuracy or that some difference occurred fails to quantify the possibility that the result was a fluke.

\subsection{Scope}
\label{sec:scope}

We restrict the scope of our methodology to making claims that an effect exists with high likelihood as quantified by the p-value.
That is, we expect our methodology to only rarely suggest that an effect exists when one does not. 

We do not claim ``completeness'' or ``power'': we might fail to detect some use of information.  For example, Google might not serve different ads upon detecting that all the browser agents in our experiment are running from the same IP address.  Despite this limitation in our experiments, we found interesting instances of usage. 

Furthermore, we do not claim that our results generalize to all users.  To do so, we would need to a take a random sample of all users, their IP addresses, browsers, and behaviors, which is prohibitively expensive. We cannot generalize our results if for example, instead of turning off some usage upon detecting our experiments, Google  turns it on. While our experiments would detect this usage, it might not be experienced by normal users.  However, it would be odd if Google purposefully performs questionable behaviors only with those attempting to find it.

While we use webpages associated with various interests to simulate users with those interests, we cannot establish that having the interest itself caused the ads to change.  It is possible that other features of the visited webpages causes change - a form of confounding called ``profile contamination''~\cite{barford14www}, since the pages cover other topics as well.  Nevertheless, we have determined that visiting webpages associated with the interest does result in seeing a change, which should give pause to users visiting webpages associated with sensitive interests.

Lastly, we do not attempt to determine how the information was used.  It could have been used by Google directly for targeting or it could have been used by advertisers to place their bids.  We cannot assign blame.  We hope future work will shed light on these issues, but given that we cannot observe the interactions between Google and advertisers, we are unsure whether it can be done.

\section{AdFisher}
\label{sec:tool}

In this section, we describe AdFisher, a tool implementing our methodology.
AdFisher makes it easy to run experiments using the above methodology for a set of treatments, measurements, and classifiers (test statistics) we have implemented.
AdFisher is also extensible allowing the experimenter to implement additional treatments, measurements, or test statistics.  
For example, an experimenter interested in studying a different online platform only needs to add code to perform actions and collect measurements on that platform.
They need not modify methods that randomize the treatments, carry out the experiment, or perform the data analysis.

To simulate a new person on the network, AdFisher creates each agent from a fresh browser instance with no browsing history, cookies, or other personalization.
AdFisher randomly assigns each agent to a group and applies the appropriate treatment, such as having the browser visit webpages.
Next, AdFisher makes measurements of the agent, such as collecting the ads shown to the browser upon visiting another webpage.
All of the agents within a block execute and finish the treatments before moving on to collect the measurements to remove time as a factor.
AdFisher runs all the agents on the same machine to prevent differences based on location, IP address, operating system, or other machine specific differences between agents.

Next, we detail the particular treatments, measurements, and test statistics that we have implemented in AdFisher.
We also discuss how AdFisher aids an experimenter in understanding the results.

\paragraph{Treatments}\label{sec:treat}
A treatment specifies what actions are to be performed by a browser agent.
AdFisher automatically applies treatments assigned to each agent.  
Typically, these treatments involve invoking the Selenium WebDriver to make the agent interact with webpages.

AdFisher makes it easy to carry out common treatments %
by providing ready-made implementations.
The simplest stock treatments we provide set interests, gender, and age range in Google's Ad Settings.  %
Another stock treatment is to visit a list of webpages stored on a file.  

To make it easy to see whether websites associated with a particular interest causes a change in behavior, we have provided the ability to create lists of webpages associated with a category on Alexa.  For each category, Alexa tracks the top websites sorted according to their traffic rank measure (a combination of the number of users and page views)~\cite{alexa}.
The experimenter can use AdFisher to download the URLs of the top webpages Alexa associates with an interest.  By default, it downloads the top  $100$ URLs. A treatment can then specify that agents visit this list of websites. %
While these treatments do not correspond directly to having such an interest, it allows us to study how Google responds to people visiting webpages associated with those interests.

Often in our experiments, we compared the effects of a certain treatment applied to the experimental group against the \emph{null treatment} applied to the control group.  Under the null treatment, agents do nothing while agents under a different treatment complete their respective treatment phase.

\paragraph{Measurements}\label{sec:col}
AdFisher can currently measure the values set in Google's Ad Settings page and the ads shown to the agents after the treatments.
It comes with stock functionality for collecting and analyzing text ads.  Experimenters can add methods for image, video, and flash ads.%

To find a reasonable website for ad collection, we looked to news sites since they generally show many ads.
Among the top $20$ news websites on \url{alexa.com}, only five displayed text ads served by Google:
\url{theguardian.com/us}, \url{timesofindia.indiatimes.com}, \url{bbc.com/news}, \url{reuters.com/news/us} and \url{bloomberg.com}. AdFisher comes with the built-in functionality to collect ads from any of these websites. %
One can also specify for how many reloads ads are to collected (default $10$), or how long to wait between successive reloads (default $5$s). %
For each page reload, AdFisher parses the page to find the ads shown by Google and stores the ads. The experimenter can add parsers to collect ads from other websites.

We run most of our experiments on Times of India as it serves the most (five) text ads per page reload. We repeat some experiments on the Guardian (three ads per reload) to demonstrate that our results are not specific to one site. %

\paragraph{Classification}\label{sec:MLclass}
While the experimenter can provide AdFisher with a test statistic to use on the collected data, AdFisher is also capable of automatically selecting a test statistic using machine learning. It splits the entire data set into training and testing subsets, and examines a training subset of the collected measurements to select a classifier that distinguishes between the measurements taken from each group.
From the point of view of machine learning, the set of ads collected by an agent corresponds to an \emph{instance} of the concept the classifier is attempting to learn.

Machine learning algorithms operate over sets of \emph{features}.  AdFisher has functions for converting the text ads seen by an agent into three different feature sets.  The \emph{URL feature set} consists of the URLs displayed by the ads (or occasionally some other text if the ad displays it where URLs normally go).  Under this feature set, the feature vector representing an agent's data has a value of $n$ in the $i$th entry iff the agent received $n$ ads that display the $i$th URL where the order is fixed but arbitrary.

The \emph{URL+Title feature set} looks at both the displayed URL and the title of the ad jointly.  It represents an agent's data as a vector where the $i$th entry is $n$ iff the agent received $n$ ads containing the $i$th pair of a URL and title.

The third feature set AdFisher has implemented is the \emph{word feature set}.  This set is based on word stems, the main part of the word with suffixes such as ``ed'' or ``ing'' removed in a manner similar to the work of Balebako et al.~\cite{balebako12w2sp}.  Each word stem that appeared in an ad is assigned a unique id.  The $i$th entry in the feature vector is the number of times that words with the $i$th stem appeared in the agent's ads.

We explored a variety of classification algorithms
provided by the scikit-learn library~\cite{scikit-learn}.
We found that logistic regression with an L2 penalty over the URL+title feature set consistently performed well compared to the others.  %
At its core, logistic regression predicts a class given a feature vector by multiplying each of the entries of the vector by its own weighting coefficient (e.g.,~\cite{bishop06pattern}).  It then takes a the sum of all these products.  If the sum is positive, it predicts one class; if negative, it predicts the other.  

While using logistic regression, the training stage consists of selecting the coefficients assigned to each feature to predict the training data.
Selecting coefficients requires balancing the training-accuracy of the model with avoiding overfitting the data with an overly complex model.  
We apply $10$-fold cross-validation on the training data to select the regularization parameter of the logistic regression classifier.
By default, AdFisher splits the data into training and test sets by using the last $10\%$ of the data collected for testing.

\paragraph{Explanations}%
To explain how the learned classifier distinguished between the groups, we explored several methods.  We found the most informative to be the model produced by the classifier itself.  %
Recall that logistic regression weighted the various features of the instances with coefficients reflecting how predictive they are of each group.  Thus, with the URL+title feature set, examining the features with the most extreme coefficients identifies the URL+title pair most used to predict the group to which agents receiving an ad with that URL+title belongs.

We also explored using simple metrics for providing explanations, like ads with the highest frequency in each group. However, some generic ads gets served in huge numbers to both groups. We also looked at the proportion of times an ad was served to agents in one group to the total number of times observed by all groups.  However, this did not provide much insight since the proportion typically reached its maximum value of $1.0$ from ads that only appeared once.
Another choice we explored was to compute the difference in the number of times an ad appears between the groups.  However, this metric is also highly influenced by how common the ad is across all groups.  %

\section{Experiments}
\label{sec:expr}

In this section, we discuss experiments that we carried out using AdFisher.  
In total, we ran 21 experiments, each of which created its own testing data sets using independent random assignments of treatments to agents.  We analyze each test data set only once and report the results of each experiment separately. Thus, we do not test multiple hypotheses on any of our test data sets ensuring that the probability of false positives (p-value) are independent with the exception of our analyses for ad choice.  In that case, we apply a Bonferroni correction.

Each experiment examines one of the properties of interest from Table~\ref{tbl:properties}. %
We found violations of nondiscrimination and data transparency and cases of compliance with effectful and ad choice.
Since these summaries each depend upon more than one experiment, they are the composite of multiple hypotheses.  
To prevent false positives for these summaries, for each property, we report p-values adjusted by the number of experiments used to explore that property.  We use the Holm-Bonferroni method for our adjustments, which is uniformly more powerful than the commonly used Bonferroni correction~\cite{holm79scandinavian}.  This method orders the component hypotheses by their unadjusted p-values applying a different correction to each until reaching a hypothesis whose adjusted value is too large to reject.  This hypothesis and all remaining hypotheses are rejected regardless of their p-values.  Appendix~\ref{app:correction} provides details.
Table~\ref{tbl:results} summarizes the results.
\begin{tablewide}
\begin{tab}{lllllrrl}
Property & Treatment & Other Actions & Source & When & Length (hrs) & \# ads & Result \\
\midrule
Nondiscrimination %
& Gender & - & TOI & May & $10$ & $40,400$ & Inconclusive\\
& Gender & Jobs & TOI & May &  $45$ & $43,393$ & Violation\\
& Gender & Jobs & TOI & July & $39$  & $35,032$ & Inconclusive\\
& Gender & Jobs & Guardian & July & $53$ & $22,596$ & Inconclusive\\
& Gender & Jobs \& Top 10 & TOI & July & $58$ & $28,738$ & Inconclusive\\
\midrule
Data use transparency %
& Substance abuse & - & TOI & May & $37$ & $42,624$ & Violation\\
& Substance abuse & - & TOI & July & $41$ & $34,408$  & Violation\\
& Substance abuse & - & Guardian & July & $51$ & $19,848$   & Violation\\
& Substance abuse & Top 10 & TOI & July & $54$ & $32,541$   & Violation\\
& Disability & - & TOI & May & $44$ & $43,136$   & Violation\\
& Mental disorder & - & TOI & May & $35$ & $44,560$  & Inconclusive\\
& Infertility & - & TOI & May & $42$ & $44,982$  & Inconclusive\\
& Adult websites & - & TOI & May & $57$ & $35,430$  & Inconclusive\\
\midrule
Effectful choice  %
& Opting out & - & TOI & May & $9$ & $18,085$  & Compliance\\
& Dating interest & - & TOI & May & $12$ & $35,737$  &  Compliance\\
& Dating interest & - & TOI & July & $17$ & $22,913$  &  Inconclusive\\
& Weight loss interest & - & TOI & May & 15 & $31,275$  &  Compliance\\
& Weight loss interest & - & TOI & July & 15 & $27,238$  &  Inconclusive\\
\midrule
Ad choice 
& Dating interest  & - & TOI & July & 1 & $1,946$   &  Compliance\\
& Weight loss interest  & - & TOI & July & 1 & $2,862$   &  Inconclusive\\
& Weight loss interest  & - & TOI & July & 1 & $3,281$   &   Inconclusive\\
\end{tab}
\caption{%
Summary of our experimental results. %
Ads are collected from the Times of India (TOI) or the Guardian. %
We report how long each experiment took, how many ads were collected for it, and what result we concluded.
}
\label{tbl:results}
\end{tablewide}

\newcommand{\topspacesub}{\vspace{0ex}}
\newcommand{\midspacesub}{\vspace{-2ex}}
\newcommand{\botspacesub}{}

\subsection{Nondiscrimination}
\label{sec:gender}

We use AdFisher to demonstrate a violation in the nondiscrimination property. %
If AdFisher finds a statistically significant difference in how Google treats two experimental groups, one consisting of members having a protected attribute and one whose members do not, then the experimenter has strong evidence that Google discriminates on that attribute.
In particular, we use AdFisher's ability to automatically select a test statistic to check for possible differences to test the null hypothesis that the two experimental groups have no differences in the ads they receive.

As mentioned before, it is difficult to send a clear signal about any attribute by visiting related webpages since they may have content related to other attributes. The only way to send a clear signal is via Ad Settings. Thus, we focus on attributes that can be set on the Ad Settings page.  
In a series of experiments, we set the gender of one group to female and the other to male.  
In one of the experiments, the agents went straight to collecting ads; in the others, they simulated an interest in jobs.  %
In all but one experiment, they collected ads from the Times of India (TOI); in the exception, they collected ads from the Guardian.
In one experiment, they also visited the top 10 websites for the U.S.\ according to \url{alexa.com} to fill out their interests.\ftnote{\url{http://www.alexa.com/topsites/countries/US}} %
Table~\ref{tbl:disc} summarizes results from these experiments.
\begin{tablewide}
\begin{tab}%
            {lllccccll} 
\multirow{2}{*}{Treatment} & \multirow{2}{*}{Other visits} & \multirow{2}{*}{Measurement} & \multirow{2}{*}{Blocks} & \multicolumn{2}{c}{\# ads (\# unique ads)} & \multirow{2}{*}{Accuracy} & \multirow{2}{*}{$\begin{array}{l}\text{Unadj.}\\[\tightlinespace]\text{p-value}\end{array}$} & \multirow{2}{*}{$\begin{array}{l}\text{Adj.}\\[\tightlinespace]\text{p-value}\end{array}$} \\
\cline{5-6}
& & & & female & male & &  \\
\midrule
Gender & Jobs & TOI, May & $100$ & $21,766$ $(545)$ & $21,627$ $(533)$ &  $93\%$ & $0.0000053$ & $0.0000265^*$\\
Gender & Jobs & Guardian, July & $100$ & $11,366$ $(410)$ & $11,230$ $(408)$ & $57\%$ & $0.12$ & $0.48$\\
Gender & Jobs \& Top 10 & TOI, July & $100$ & $14,507$ $(461)$ & $14,231$ $(518)$ & $56\%$ & $0.14$ & n/a\\
Gender & Jobs & TOI, July & $100$ & $17,019$ $(673)$ & $18,013$ $(690)$ &  $55\%$ & $0.20$ & n/a\\
Gender & - & TOI, May & $100$ &  $20,137$ $(603)$ & $20,263$ $(630)$ &  $48\%$ & $0.77$ & n/a\\
\end{tab}
\caption{Results from the discrimination experiments sorted by unadjusted p-value. TOI stands for Times of India. $\mbox{}^*$ denotes statistically significant results under the Holm-Bonferroni method.}
\label{tbl:disc}
\end{tablewide}

AdFisher found a statistically significant difference in the ads for male and female agents that simulated an interest in jobs in May, 2014.
It also found evidence of discrimination in the nature of the effect.  
In particular, 
it found that females received fewer instances of an ad encouraging the taking of high paying jobs than males.  
AdFisher did not find any statistically significant differences among the agents that did not visit the job-related pages or those operating in July, 2014.
We detail the experiment finding a violation before discussing why we think the other experiments did not result in significant results.

\paragraph{Gender and Jobs}
In this experiment, we examine how changing the gender demographic on Google Ad Settings affects the ads served and interests inferred for agents browsing employment related websites.
We set up AdFisher to have the agents in one group visit the Google Ad Settings page and set the gender bit to female while agents in the other group set theirs to male.
All the agents then visited the top $100$ websites listed under the Employment category of Alexa \ftnote{\url{http://www.alexa.com/topsites/category/Top/Business/Employment}}.
The agents then collect ads from Times of India.

AdFisher ran $100$ blocks of $10$ agents each. (We used blocks of size $10$ in all our experiments.)
AdFisher used the ads of $900$ agents ($450$ from each group) for training a classifier using the URL+title feature set, and used the remaining $100$ agents' ads for testing.  
The learned classifier attained a test-accuracy of $93\%$, suggesting that Google did in fact treat the genders differently.
To test whether this response was statistically significant, AdFisher computed a p-value by running the permutation test on a million randomly selected block-respecting permutations of the data.
The significance test yielded an adjusted p-value of $< 0.00005$. %

We then examined the model learned by AdFisher to explain the nature of the difference.
Table~\ref{tab:gen+jobs-featsel} shows the five URL+title pairs that the model identifies as the strongest indicators of being from the female or male group.
How ads for identifying the two groups differ is concerning. 
The two URL+title pairs with the highest coefficients for indicating a male were for a career coaching service for ``$\$200$k+'' executive positions. 
Google showed the ads $1852$ times to the male group but just $318$ times to the female group.
The top two URL+title pairs for the female group was for a generic job posting service and for an auto dealer.
\begin{tablewide}
\begin{tab}{llrrrcrr}
\multirow{2}{*}{Title} & \multirow{2}{*}{URL} & \multirow{2}{*}{Coefficient} & \multicolumn{2}{c}{appears in agents} && \multicolumn{2}{c}{total appearances} \\
\cline{4-5} \cline{7-8}
 &  &  & female & male && female & male\\
\midrule
\multicolumn{8}{c}{Top ads for identifying the simulated female group} \\
\midruleheaderbottom
Jobs (Hiring Now)  & \url{ www.jobsinyourarea.co } & $ 0.34 $ & $ 6 $ & $ 3 $ & & $ 45 $ & $ 8 $ \\
4Runner Parts  Service  & \url{ www.westernpatoyotaservice.com } & $ 0.281 $ & $ 6 $ & $ 2 $ & & $ 36 $ & $ 5 $ \\
Criminal Justice Program  & \url{ www3.mc3.edu/Criminal+Justice } & $ 0.247 $ & $ 5 $ & $ 1 $ & & $ 29 $ & $ 1 $ \\
Goodwill - Hiring  & \url{ goodwill.careerboutique.com } & $ 0.22 $ & $ 45 $ & $ 15 $ & & $ 121 $ & $ 39 $ \\
UMUC Cyber Training  & \url{ www.umuc.edu/cybersecuritytraining } & $ 0.199 $ & $ 19 $ & $ 17 $ & & $ 38 $ & $ 30 $ \\
\midrule
\multicolumn{8}{c}{Top ads for identifying agents in the simulated male group}\\
\midruleheaderbottom
\$200k+ Jobs - Execs Only  & \url{ careerchange.com } & $ -0.704 $ & $ 60 $ & $ 402 $ & & $ 311 $ & $ 1816 $ \\
Find Next \$200k+ Job  & \url{ careerchange.com } & $ -0.262 $ & $ 2 $ & $ 11 $ & & $ 7 $ & $ 36 $ \\
Become a Youth Counselor  & \url{ www.youthcounseling.degreeleap.com } & $ -0.253 $ & $ 0 $ & $ 45 $ & & $ 0 $ & $ 310 $ \\
CDL-A OTR Trucking Jobs  & \url{ www.tadrivers.com/OTRJobs } & $ -0.149 $ & $ 0 $ & $ 1 $ & & $ 0 $ & $ 8 $ \\
Free Resume Templates  & \url{ resume-templates.resume-now.com } & $ -0.149 $ & $ 3 $ & $ 1 $ & & $ 8 $ & $ 10 $ \\
\end{tab}
\onlyarxiv{\midspacesub}
\caption{Top URL+titles for the gender and jobs experiment on the Times of India in May.}
\label{tab:gen+jobs-featsel}
\end{tablewide}

The found discrimination in this experiment was predominately from a pair of job-related ads for the same service making the finding highly sensitive to changes in the serving of these ads.
A closer examination of the ads from the same experimental setup ran in July, 2014, showed that the frequency of these ads reduced from $2170$ to just $48$, with one of the ads completely disappearing (Table~\ref{tbl:gen+jobs-july-featsel}).
These $48$ ads were only shown to males, continuing the pattern of discrimination.
This pattern was recognized by the machine learning algorithm, which selected the ad as the second most useful for identifying males.  %
However, they were too infrequent to establish statistical significance.
A longer running experiment with more blocks might have succeeded.

\begin{tablewide}
\begin{tab}{@{}llrrrcrr@{}}
\multirow{2}{*}{Title} & \multirow{2}{*}{URL} & \multirow{2}{*}{Coefficient} & \multicolumn{2}{c}{appears in agents} && \multicolumn{2}{c}{total appearances} \\
\cline{4-5} \cline{7-8}
 &  &  & female & male && female & male\\
\midrule
\multicolumn{8}{c}{Top ads for identifying the simulated male group} \\
\midruleheaderbottom
Truck Driving Jobs  & \url{ www.bestpayingdriverjobs.com } & $ 0.492 $ & $0$ & $ 15 $ & & $0$ & $ 33 $\\
\$200k+ Jobs - Execs Only  & \url{ careerchange.com } & $ 0.383 $      & $0$ & $ 15 $ & & $0$ & $ 48 $ \\
Aircraft Technician Program  & \url{ pia.edu } & $ 0.292 $ & $0$ & $ 6 $ & & $0$ & $ 14 $ \\
Paid CDL Training  & \url{ pamtransport.greatcdltraining.com } & $ 0.235 $ & $0$ & $ 5 $ & & $0$ & $ 13 $\\
Unique Bridal Necklaces  & \url{ margaretelizabeth.com/Bridal } & $ 0.234 $ & $0$ & $ 5 $ & & $0$ & $ 19 $\\
\midrule
\multicolumn{8}{c}{Top ads for identifying agents in the simulated female group}\\
\midruleheaderbottom
Business Loans for Women  & \url{ topsbaloans.com } & $ -0.334 $ & $ 13 $ & $ 1 $ & & $ 70 $ & $ 1 $ \\
Post Your Classified Ad  & \url{ indeed.com/Post-Jobs } & $ -0.267 $ & $ 20 $ & $ 16 $ & & $ 56 $ & $ 24 $ \\
American Laser Skincare  & \url{ americanlaser.com } & $ -0.243 $ & $ 8 $ & $ 5 $ & & $ 14 $ & $ 8 $ \\
Dedicated Drivers Needed  & \url{ warrentransport.com } & $ -0.224 $ & $ 3 $ & $ 0 $ & & $ 14 $ & $ 0 $ \\
Earn Your Nursing Degree  & \url{ nursing-colleges.courseadvisor.com } & $ -0.219 $ & $ 11 $ & $ 3 $ & & $ 31 $ & $ 10 $ \\
\end{tab}
\onlyarxiv{\midspacesub}
\caption{Top URL+titles for the gender and jobs experiment (July).} %
\label{tbl:gen+jobs-july-featsel}
\end{tablewide}

\subsection{Transparency}
\label{sec:subs}

AdFisher can demonstrate violations of individual data use transparency. 
AdFisher tests the null hypothesis that two groups of agents with the same ad settings receives ads from the same distribution despite being subjected to different experimental treatments.
Rejecting the null hypothesis implies that some difference exists in the ads that is not documented by the ad settings.

In particular, we ran a series of experiments to examine how much transparency Google's Ad Settings provided.  We checked whether visiting webpages associated with some interest could cause a change in the ads shown that is not reflected in the settings.

We ran such experiments for five interests: substance abuse, disabilities, infertility\ftnote{\url{http://www.alexa.com/topsites/category/Top/Health/Reproductive_Health/Infertility}},
mental disorders\ftnote{\url{http://www.alexa.com/topsites/category/Top/Health/Mental_Health/Disorders}}, and
adult websites\ftnote{\url{http://www.alexa.com/topsites/category/Top/Adult}}. 
Results from statistical analysis of these experiments are shown in Table~\ref{tbl:opacity}. %
\begin{tablewide}
\begin{tab}{lllcccll} 
\multirow{2}{*}{Treatment} & \multirow{2}{*}{Other visits} & \multirow{2}{*}{Measurement} 
& \multicolumn{2}{c}{\# ads (\# unique ads)} & \multirow{2}{*}{Accuracy} & \multirow{2}{*}{$\begin{array}{l}\text{Unadj.}\\[\tightlinespace]\text{p-value}\end{array}$} & \multirow{2}{*}{$\begin{array}{l}\text{Adj.}\\[\tightlinespace]\text{p-value}\end{array}$} \\
\cline{4-5}
& & & experimental & control & &  \\
\midrule
Substance abuse & - & TOI, May & $20,420$ $(427)$ & $22,204$ $(530)$  & $81\%$ & $0.0000053$ & $0.0000424^*$\\
Substance abuse & - & TOI, July & $16,206$ $(653)$ & $18,202$ $(814)$  & $98\%$ & $0.0000053$ & $0.0000371^*$\\
Substance abuse & Top 10 & TOI, July & $15,713$ $(603)$ & $16,828$ $(679)$ & $65\%$ & $0.0000053$ & $0.0000318^*$\\

Disability & - & TOI, May & $19,787$ $(546)$ & $23,349$ $(684)$ &  $75\%$ & $0.0000053$ & $0.0000265^*$\\
Substance abuse & - & Guardian, July & $8,359$ $(242)$ & $11,489$ $(319)$ & $62\%$ & $0.0075$ & $0.03^*$\\
Mental disorder & - & TOI, May & $22,303$ $(407)$ &  $22,257$ $(465)$ & $59\%$ & $0.053$ & $0.159$\\
Infertility & - & TOI, May & $22,438$ $(605)$ &  $22,544$ $(625)$ & $57\%$ & $0.11$ & n/a\\
Adult websites & - & TOI, May & $17,670$ $(602)$ &  $17,760$ $(580)$ & $52\%$ & $0.42$ & n/a\\
\end{tab}
\caption{Results from transparency experiments. TOI stands for Times of India.  Every experiment for this property ran with $100$ blocks.  $\mbox{}^*$ denotes statistically significant results under the Holm-Bonferroni method.}
\label{tbl:opacity}
\end{tablewide}

We examined the interests found in the settings for the two cases where we found a statistically significant difference in ads, substance abuse and disability.
We found that settings did not change at all for substance abuse and changed in an unexpected manner for disabilities.
Thus, we detail these two experiments below.

\paragraph{Substance Abuse}
We were interested in whether Google's outputs would change in response to visiting webpages associated with substance abuse, a highly sensitive topic.  Thus, we ran an experiment in which the experimental group visited such websites while the control group idled.  Then, we collected the Ad Settings and the Google ads shown to the agents at the Times of India.
For the webpages associated with substance abuse, we used the top $100$ websites on the Alexa list for substance abuse\ftnote{\url{http://www.alexa.com/topsites/category/Top/Health/Addictions/Substance_Abuse}}.

AdFisher ran $100$ blocks of $10$ agents each. 
At the end of visiting the webpages associated with substance abuse, none of the $500$ agents in the experimental group had interests listed on their Ad Settings pages.  (None of the agents in the control group did either since the settings start out empty.)
If one expects the Ad Settings page to reflect all learned inferences, then he would not anticipate ads relevant to those website visits given the lack of interests listed.

However, the ads collected from the Times of India told a different story.
The learned classifier attained a test-accuracy of $81\%$, suggesting that Google did in fact respond to the page visits.  
Indeed, using the permutation test, AdFisher found an adjusted p-value of $< 0.00005$. %
Thus, we conclude that the differences are statistically significant: Google's ads changed in response to visiting the webpages associated with substance abuse.
Despite this change being significant, the Ad Settings pages provided no hint of its existence: the transparency tool is opaque!

\begin{figure}
\onlyarxiv{\vspace{2ex}}
\centerline{\includegraphics[width=16cm]{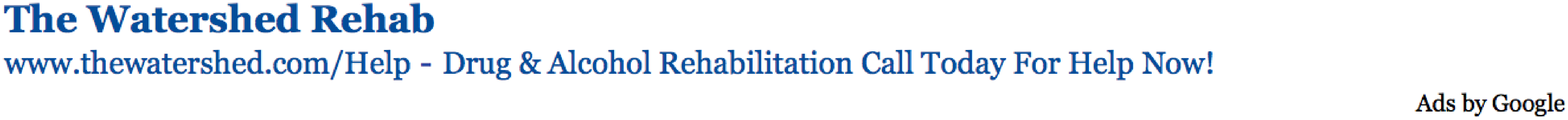}}
\onlyarxiv{\vspace{-3ex}}
\caption{Screenshot of an ad with the top URL+title for identifying agents that visited webpages associated with substance abuse}
\label{fig:rehab}
\end{figure}

We looked at the URL+title pairs with the highest coefficients for identifying the experimental group that visited the websites related to substance abuse.  
Table~\ref{tab:subs-featsel} provides information on coefficients and URL+titles learned.
The three highest were for ``Watershed Rehab''.  The top two had URLs for this drug and alcohol rehab center.  The third lacked a URL and had other text in its place.
Figure~\ref{fig:rehab} shows one of Watershed's ads.
\begin{tablewide}
\begin{tab}{llrrrcrr}
\multirow{2}{*}{Title} & \multirow{2}{*}{URL} & \multirow{2}{*}{Coefficient} & \multicolumn{2}{c}{appears in agents} && \multicolumn{2}{c}{total appearances} \\
\cline{4-5} \cline{7-8}
 &  &  & control & experi. && control & experi.\\
\midrule
\multicolumn{8}{c}{Top ads for identifying agents in the experimental group (visited websites associated with substance abuse)}\\
\midruleheaderbottom
The Watershed Rehab & \url{www.thewatershed.com/Help} & $-0.888$ &  0  &  280  &&  0 &  2276  \\
Watershed Rehab & \url{www.thewatershed.com/Rehab} &  $-0.670$ &  0  &  51  &&  0 &  362  \\
The Watershed Rehab & Ads by Google & $-0.463$  & 0  &  258  &&  0 &  771  \\
Veteran Home Loans & \url{www.vamortgagecenter.com} & $-0.414$ & 13  &  15  &&  22 &  33  \\
CAD Paper Rolls & \url{paper-roll.net/Cad-Paper} &  $-0.405$ &  0  &  4  &&  0 &  21  \\
\midrule
\multicolumn{8}{c}{Top ads for identifying agents in control group} \\
\midruleheaderbottom
Alluria Alert & \url{www.bestbeautybrand.com} & $0.489$ &  2  &  0  &&  9 &  0  \\
Best Dividend Stocks & \url{dividends.wyattresearch.com} &  $0.431$ &  20  &  10  &&  54 &  24  \\
10 Stocks to Hold Forever & \url{www.streetauthority.com} & $0.428$ &  51  &  44  &&  118 &  76  \\
Delivery Drivers Wanted & \url{get.lyft.com/drive} & $0.362$ &  22  &  6  &&  54 &  14  \\
VA Home Loans Start Here & \url{www.vamortgagecenter.com} & $0.354$ &  23  &  6  &&  41 &  9  \\
\end{tab}
\onlyarxiv{\vspace{-1ex}}
\caption{Top URL+titles for substance abuse experiment on the Times of India in May. }
\label{tab:subs-featsel}
\end{tablewide}
The experimental group saw these ads a total of $3309$ times ($16\%$ of the ads); the control group never saw any of them nor contained any ads with the word ``rehab'' or ``rehabilitation''.
None of the top five URL+title pairs for identifying the control group had any discernible relationship with rehab or substance abuse.

These results remain robust across variations on this design with statistical significance in three variations.
In July, we repeated the aforementioned experiment, conducted a variation using the Guardian instead of the Times of India, and conducted a variation that involved all the agents first visiting the top 10 websites in addition to the agents in the experimental group also visiting the substance abuse websites.  In each of these experiments, there were two of Watershed ads that were the top two ads for identifying the agents that visited the substance abuse websites (Tables~\ref{tab:subs-featsel-july-toi}, \ref{tab:subs-featsel-july-g}, and~\ref{tab:subs-featsel-july-toi-top10}).
\begin{tablewide}
\onlyarxiv{\topspacesub}
\begin{tab}{@{}llrrrcrr@{}}
\multirow{2}{*}{Title} & \multirow{2}{*}{URL} & \multirow{2}{*}{Coefficient} & \multicolumn{2}{c}{appears in agents} && \multicolumn{2}{c}{total appearances} \\
\cline{4-5} \cline{7-8}
 &  &  & control & experi. && control & experi.\\
\midrule
\multicolumn{8}{c}{Top ads for identifying agents in the experimental group (visited websites associated with substance abuse)}\\
\midruleheaderbottom
The Watershed Rehab  & \url{ www.thewatershed.com/Help } & $ -0.958 $ & $ 0 $ & $ 310 $ & & $ 0 $ & $ 1674 $ \\
The Watershed Rehab  &  Ads by Google  & $ -0.532 $ & $ 0 $ & $ 259 $ & & $ 0 $ & $ 721 $ \\
2014 Diabetes Risk Survey  & \url{ prediabetescenters.com } & $ -0.195 $ & $ 6 $ & $ 7 $ & & $ 12 $ & $ 23 $ \\
Honda CR-V Clearance 2014  & \url{ honda-clearance-sale.autosite.com } & $ -0.191 $ & $ 9 $ & $ 8 $ & & $ 23 $ & $ 41 $ \\
Considering an eMBA?  & \url{ gsb.stanford.edu/eMBAAlternative } & $ -0.172 $ & $ 4 $ & $ 8 $ & & $ 23 $ & $ 51 $ \\
\midrule
\multicolumn{8}{c}{Top ads for identifying agents in control group} \\
\midruleheaderbottom
Best Dividend Stocks  & \url{ dividends.wyattresearch.com } & $ 0.314 $ & $ 100 $ & $ 80 $ & & $ 552 $ & $ 340 $ \\
Luxury Villas in Gurgaon  & \url{ tatahousing.in/Arabella_EnquireNow } & $ 0.201 $ & $ 48 $ & $ 41 $ & & $ 176 $ & $ 146 $ \\
Apply for Discover{\tiny$\circledR$} it  & \url{ www.discovercard.com } & $ 0.183 $ & $ 14 $ & $ 2 $ & & $ 43 $ & $ 13 $ \\
Man Cheats Credit Score  & \url{www.thecreditsolutionprogram.com} & $ 0.163 $ & $ 36 $ & $ 27 $ & & $ 93 $ & $ 66 $ \\
Diabetes Signs  Symptoms  & \url{ prediabetescenters.com/Symptoms } & $ 0.153 $ & $ 18 $ & $ 7 $ & & $ 72 $ & $ 16 $ \\
\end{tab}
\onlyarxiv{\midspacesub}
\caption{Top URL+titles for substance abuse experiment on the Times of India in July.}
\label{tab:subs-featsel-july-toi}
\onlyarxiv{\botspacesub}
\end{tablewide}
\begin{tablewide}
\begin{tab}{@{}llrrrcrr@{}}
\multirow{2}{*}{Title} & \multirow{2}{*}{URL} & \multirow{2}{*}{Coefficient} & \multicolumn{2}{c}{appears in agents} && \multicolumn{2}{c}{total appearances} \\
\cline{4-5} \cline{7-8}
 &  &  & control & experi. && control & experi.\\
\midrule
\multicolumn{8}{c}{Top ads for identifying agents in the experimental group (visited websites associated with substance abuse)}\\
\midruleheaderbottom
The Watershed Rehab  & \url{ www.thewatershed.com/Help } & $ -0.626 $ & $ 0 $ & $ 231 $ & & $ 0 $ & $ 1847 $ \\
Watershed Rehab  & \url{ www.thewatershed.com/Rehab } & $ -0.207 $ & $ 0 $ & $ 12 $ & & $ 0 $ & $ 52 $ \\
Generator Sets  Parts  & \url{ www.mtspowerproducts.com } & $ -0.152 $ & $ 0 $ & $ 1 $ & & $ 0 $ & $ 10 $ \\
The AntiChrist: Free Book  & \url{ voiceofelijah.org/Rapture } & $ -0.148 $ & $ 1 $ & $ 11 $ & & $ 6 $ & $ 17 $ \\
Israel at War  & \url{ www.joelrosenberg.com } & $ -0.148 $ & $ 1 $ & $ 4 $ & & $ 1 $ & $ 19 $ \\
\midrule
\multicolumn{8}{c}{Top ads for identifying agents in control group} \\
\midruleheaderbottom
The Sound of Dear Voices  & \url{ www.telephonebangladesh.com } & $ 0.243 $ & $ 21 $ & $ 3 $ & & $ 59 $ & $ 5 $ \\
5-15 Day Ireland Vacation  & \url{ www.exploringvacations.com/Ireland } & $ 0.233 $ & $ 21 $ & $ 7 $ & & $ 109 $ & $ 18 $ \\
\#1 Best Selling Blocker  & \url{ plugnblock.com/Sentry-Call-Blocker } & $ 0.207 $ & $ 37 $ & $ 22 $ & & $ 87 $ & $ 33 $ \\
Dow Average Over 17,000  & \url{ economyandmarkets.com } & $ 0.204 $ & $ 14 $ & $ 1 $ & & $ 34 $ & $ 2 $ \\
Block Annoying Phone Call  & \url{ plugnblock.com/Sentry-Call-Blocker } & $ 0.176 $ & $ 27 $ & $ 18 $ & & $ 62 $ & $ 25 $ \\
\end{tab}
\onlyarxiv{\midspacesub}
\caption{Top URL+titles for substance abuse experiment on the Guardian in July} %
\label{tab:subs-featsel-july-g}
\onlyarxiv{\botspacesub}
\end{tablewide}
\begin{tablewide}
\begin{tab}{@{}llrrrcrr@{}}
\multirow{2}{*}{Title} & \multirow{2}{*}{URL} & \multirow{2}{*}{Coefficient} & \multicolumn{2}{c}{appears in agents} && \multicolumn{2}{c}{total appearances} \\
\cline{4-5} \cline{7-8}
 &  &  & control & experi. && control & experi.\\
\midrule
\multicolumn{8}{c}{Top ads for identifying agents in the experimental group (visited websites associated with substance abuse)}\\
\midruleheaderbottom
The Watershed Rehab  & \url{ www.thewatershed.com/Help } & $ -1.942 $ & $ 0 $ & $ 213 $ & & $ 0 $ & $ 1211 $ \\
The Watershed Rehab  & \url{ Ads by Google } & $ -1.41 $ & $ 0 $ & $ 174 $ & & $ 0 $ & $ 413 $ \\
3 veggies keeping you fat  & \url{ www.beyonddiet.com } & $ -0.851 $ & $ 1 $ & $ 2 $ & & $ 1 $ & $ 8 $ \\
Flexible Jobs Available  & \url{ get.lyft.com/drive } & $ -0.794 $ & $ 2 $ & $ 4 $ & & $ 3 $ & $ 6 $ \\
Accord Clearance 2014  & \url{ honda-clearance-sale.autosite.com } & $ -0.783 $ & $ 32 $ & $ 122 $ & & $ 39 $ & $ 167 $ \\
\midrule
\multicolumn{8}{c}{Top ads for identifying agents in control group} \\
\midruleheaderbottom
Lung Cancer Symptoms  & \url{ symptomfind.com/LungCancer } & $ 0.822 $ & $ 32 $ & $ 6 $ & & $ 88 $ & $ 12 $ \\
Women's Gardening Clothes  & \url{ duluthtrading.com/Women } & $ 0.762 $ & $ 2 $ & $ 0 $ & & $ 10 $ & $ 0 $ \\
Avacor{\tiny$\circledR$} - Official Site  & \url{ www.avacor.com } & $ 0.626 $ & $ 3 $ & $ 0 $ & & $ 6 $ & $ 0 $ \\
5-15 Day Ireland Vacation  & \url{ www.exploringvacations.com/Ireland } & $ 0.62 $ & $ 14 $ & $ 3 $ & & $ 33 $ & $ 4 $ \\
Prostate Cancer Treatment  & \url{ prostrcision.com/Prostate_Cancer } & $ 0.593 $ & $ 7 $ & $ 7 $ & & $ 30 $ & $ 32 $ \\
\end{tab}
\onlyarxiv{\midspacesub}
\caption{Top URL+titles for substance abuse experiment on the Times of India in July.} %
\onlyarxiv{\botspacesub}
\label{tab:subs-featsel-july-toi-top10}
\end{tablewide}

One possible reason why Google served Watershed's ads could be \emph{remarketing}, a marketing strategy that encourages users to return to previously visited websites~\cite{google-remarketing}. The website \url{thewatershed.com} features among the top $100$ websites about substance-abuse on Alexa, and agents visiting that site may be served Watershed's ads as part of remarketing. %
However, these users cannot see any changes on Google Ad Settings despite Google having learnt some characteristic (visited \url{thewatershed.com}) about them and serving ads relevant to that characteristic. %
\begin{wrapfigure}{}{9.1cm}
\includegraphics[width = 9cm]{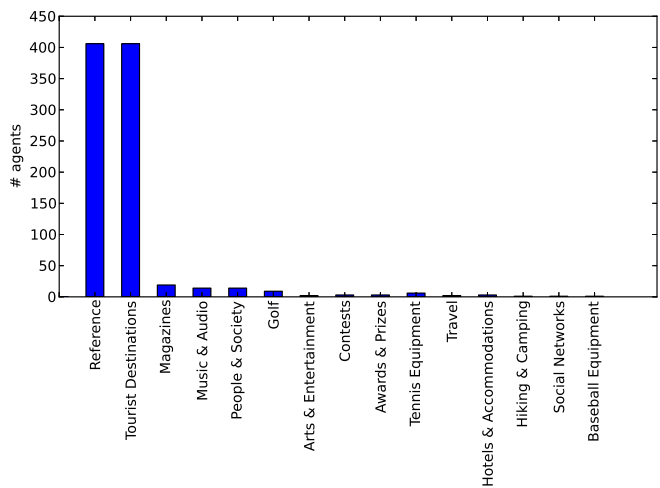}
\caption{For each interest selected for the agents that visited webpages associated with disabilities, the number of agents with that interest selected}
\label{fig:disability-settings}
\end{wrapfigure}

\paragraph{Disabilities}
This experiment was nearly identical in setup but used websites related to disabilities instead of substance abuse.
We used the top $100$ websites on Alexa on the topic.\ftnote{\url{http://www.alexa.com/topsites/category/Top/Society/Disabled}}

For this experiment, AdFisher found a classifier with a test-accuracy of $75\%$.  It found a statistically significant difference with an adjusted p-value of less than $0.00005$. %

Looking at the top ads for identifying agents that visited the webpages associated with disabilities, we see that the top two ads have the URL \url{www.abilitiesexpo.com} and the titles ``Mobility Lifter'' and ``Standing Wheelchairs''.
They were shown a total of $1076$ times to the experimental group but never to the control group.
(See Table~\ref{tab:disability-featsel}.)
Table~\ref{tab:disability-featsel} showing them for the disability experiment.
\begin{tablewide}
\begin{tab}{llrrrcrr}
\multirow{2}{*}{Title} & \multirow{2}{*}{URL} & \multirow{2}{*}{Coefficient} & \multicolumn{2}{c}{appears in agents} && \multicolumn{2}{c}{total appearances} \\
\cline{4-5} \cline{7-8}
& & & control & experi. && control & experi.\\
\midrule
\multicolumn{8}{c}{Top ads for identifying agents in the experimental group (visited websites associated with disability)}\\
\midruleheaderbottom
Mobility Lifter & \url{ www.abilitiesexpo.com } & $ -1.543 $ & $ 0 $ & $ 84 $ & & $ 0 $ & $ 568 $ \\
Standing Wheelchairs & \url{ www.abilitiesexpo.com } & $ -1.425 $ & $ 0 $ & $ 88 $ & & $ 0 $ & $ 508 $ \\
Smoking MN Healthcare & \url{ www.stillaproblem.com } & $ -1.415 $ & $ 0 $ & $ 24 $ & & $ 0 $ & $ 60 $ \\
Bike Prices & \url{ www.bikesdirect.com } & $ -1.299 $ & $ 0 $ & $ 24 $ & & $ 0 $ & $ 79 $ \\
\$19 Car Insurance - New & \url{ auto-insurance.quotelab.com/MN } & $ -1.276 $ & $ 0 $ & $ 6 $ & & $ 0 $ & $ 9 $ \\
\midrule
\multicolumn{8}{c}{Top ads for identifying agents in control group} \\
\midruleheaderbottom
Beautiful Women in Kiev & \url{ anastasiadate.com } & $ 1.304 $ & $ 190 $ & $ 46 $ & & $ 533 $ & $ 116 $ \\
Melucci DDS & \url{ Ads by Google } & $ 1.255 $ & $ 4 $ & $ 2 $ & & $ 10 $ & $ 6 $ \\
17.2\% 2013 Annuity Return & \url{ advisorworld.com/CompareAnnuities } & $ 1.189 $ & $ 30 $ & $ 5 $ & & $ 46 $ & $ 6 $ \\
3 Exercises To Never Do & \url{ homeworkoutrevolution.net } & $ 1.16 $ & $ 1 $ & $ 1 $ & & $ 3 $ & $ 1 $ \\
Find CNA Schools Near You & \url{ cna-degrees.courseadvisor.com } & $ 1.05 $ & $ 22 $ & $ 0 $ & & $ 49 $ & $ 0 $ \\
\end{tab}
\onlyarxiv{\midspacesub}
\caption{Top URL+titles for disability experiment on the Times of India in May.}
\label{tab:disability-featsel}
\end{tablewide}

This time, Google did change the settings in response to the agents visiting the websites.
Figure~\ref{fig:disability-settings} shows the interests selected for the experimental group.  (The control group, which did nothing, had no interests selected.)
None of them are directly related to disabilities suggesting that Google might have focused on other aspects of the visited pages.  Once again, we believe that the top ads were served due to remarketing, as \url{abilitiesexpo.com} was among the top $100$ websites related to disabilities. %

\subsection{Effectful Choice}

We tested whether making changes to Ad Settings has an effect on the ads seen, thereby giving the users a degree of choice over the ads.  In particular, AdFisher tests the null hypothesis that changing some ad setting has no effect on the ads.

First, we tested whether opting out of tracking actually had an effect by comparing the ads shown to agents that opted out after visiting car-related websites to ads from those that did not opt out.
We found a statistically significant difference.

We also tested whether removing interests from the settings page actually had an effect.
We set AdFisher to have both groups of agents simulate some interest. 
AdFisher then had the agents in one of the groups remove interests from Google's Ad Settings related to the induced interest.
We found statistically significant differences between the ads both groups collected from the Times of India for two induced interests: online dating and weight loss.  
Table~\ref{tbl:effectful-choice} summarizes the results.
We describe one in detail below.

\begin{tablewide}
\begin{tab}%
            {lrrrrcll} 
\multirow{2}{*}{Experiment} & \multirow{2}{*}{blocks} & \multicolumn{3}{c}{\# ads (\# unique ads)} & \multirow{2}{*}{accuracy} & \multirow{2}{*}{$\begin{array}{l}\text{Unadj.}\\[\tightlinespace]\text{p-value}\end{array}$} & \multirow{2}{*}{$\begin{array}{l}\text{Adj.}\\[\tightlinespace]\text{p-value}\end{array}$} \\
\cline{3-5}
& & removed/opt-out & keep/opt-in & total & & \\
\midrule
Opting out   & $54$  & $9,029$ $(139)$ & $9,056$ $(293)$   & $18,085$ $(366)$ & $83\%$ & $0.0000053$ & $0.0000265^*$\\
Dating (May) & $100$ & $17,975$ $(518)$ & $17,762$ $(457)$ & $35,737$ $(669)$ & $74\%$ & $0.0000053$ & $0.0000212^*$\\
Weight Loss (May) & $83$ & $15,826$ $(367)$ & $15,449$ $(427)$ & $31,275$ $(548)$ & $60\%$ & $0.041$ & $0.123$\\
Dating (July) & $90$ & $11,657$ $(727)$ & $11,256$ $(706)$ & $22,913$ $(1,014)$ & $59\%$ & $0.070$ & n/a\\
Weight Loss (July) & $100$ & $14,168$ $(917)$ & $13,070$ $(919)$ & $27,238$ $(1,323)$ & $52\%$ & $0.41$ & n/a\\
\end{tab}
\caption{Results from effectful choice experiments using the Times of India sorted by unadjusted p-value.  $\mbox{}^*$ denotes statistically significant results under the Holm-Bonferroni method.}
\label{tbl:effectful-choice}
\end{tablewide}

\paragraph{Online Dating}

We simulated an interest in online dating by visiting the website \url{www.midsummerseve.com/}, a website we choose since it sets Google's ad setting for ``Dating \& Personals'' (this site no longer affects the setting).
AdFisher then had just the agents in the experimental group remove the interest ``Dating \& Personals'' (the only one containing the keyword ``dating'').  All the agents then collected ads from the Times of India.

AdFisher found statistically significant differences between the groups with a classifier accuracy of 74\% and an adjusted p-value of $< 0.00003$.   
Furthermore, the effect appears related to the interests removed.  The top ad for identifying agents that kept the romantic interests has the title ``Are You Single?'' and the second ad's title is ``Why can't I find a date?''.
 None of the top five for the control group that removed the interests were related to dating (Table~\ref{tbl:dating-featsel}).
\begin{tablewide}
\begin{tab}{llrrrcrr}
\multirow{2}{*}{Title} & \multirow{2}{*}{URL} & \multirow{2}{*}{Coefficient} & \multicolumn{2}{c}{appears in agents} && \multicolumn{2}{c}{total appearances} \\
\cline{4-5} \cline{7-8}
 & & & kept & removed && kept & removed \\
\midrule
\multicolumn{8}{c}{Top ads for identifying the group that kept dating interests} \\
\midruleheaderbottom
Are You Single? & \url{www.zoosk.com/Dating} & $1.583$ &  367  &  33  &&  2433 &  78  \\
Top 5 Online Dating Sites & \url{www.consumer-rankings.com/Dating} & $1.109$ &  116  &  10  &&  408 &  13  \\
Why can't I find a date? & \url{www.gk2gk.com} & $0.935$ &  18  &  3  &&  51 &  5  \\
Latest Breaking News & \url{www.onlineinsider.com} & $0.624$ &  2  &  1  &&  6 &  1\\
Gorgeous Russian Ladies & \url{anastasiadate.com} & $0.620$  &  11  &  0  &&  21 &  0  \\
\midrule
\multicolumn{8}{c}{Top ads for identifying agents in the group that removed dating interests}\\
\midruleheaderbottom
Car Loans w/ Bad Credit & \url{www.car.com/Bad-Credit-Car-Loan} & $-1.113$ &  5  &  13  &&  8 &  37  \\
Individual Health Plans & \url{www.individualhealthquotes.com} &  $-0.831$ &  7  &  9  &&  21 &  46  \\
Crazy New Obama Tax & \url{www.endofamerica.com} & $-0.722$ &  19  &  31  &&  22 &  51  \\
Atrial Fibrillation Guide & \url{www.johnshopkinshealthalerts.com} &  $-0.641$ &  0  &  6  &&  0 &  25  \\
Free \$5 - \$25 Gift Cards & \url{swagbucks.com} & $-0.614$ &  4  &  11  &&  5 &  32 \\
\end{tab}
\onlyarxiv{\midspacesub}
\caption{Top URL+titles for the dating experiment on Times of India in May.}
\label{tbl:dating-featsel}
\end{tablewide}
Thus, the ad settings appear to actually give users the ability to avoid ads they might dislike or find embarrassing.
In the next set of experiments, we explicitly test for this ability.

We repeated this experiment in July, 2014, using the websites \url{relationshipsurgery.com} and \url{datemypet.com}, which also had an effect on Ad Settings, but did not find statistically significant differences.

\subsection{Ad Choice}

Whereas the other experiments tested merely for the presence of an effect, testing for ad choice requires determining whether the effect is an increase or decrease in the number of relevant ads seen.
Fortunately, since AdFisher uses a one-sided permutation test, it tests for either an increase or a decrease, but not for both simultaneously, making it usable for this purpose.
In particular, after removing an interest, we check for a decrease to test for compliance using the null hypothesis that either no change or an increase occurred, since rejecting this hypothesis would imply that a decrease in the number of related ads occurred.
To check for a violation, we test for the null hypothesis that either no change or a decrease occurred.
Due to testing two hypotheses, we use an adjustment to the p-value cutoff considered significant to avoid finding significant results simply from testing multiple hypotheses.
In particular, we use the standard Bonferroni correction, which calls for multiplying the p-value by $2$ (e.g.,~\cite{abdi07enc}).

We ran three experiments checking for ad choice.
The experiments followed the same setup as the effectful choice ones, but this time we used all the blocks for testing a given test statistic.  The test statistic counted the number of ads containing keywords.  
In the first, we again test online dating using \url{relationshipsurgery.com} and \url{datemypet.com}.
Table~\ref{tbl:ad-choice-ads} summarizes the experimental setups and Table~\ref{tbl:ad-choice-p-values} summarizes the results.
In particular, we found that removing online dating resulted in a significant decrease (p-value adjusted for all six experiments: $0.0456$) in the number of ads containing related keywords (from $109$ to $34$).  We detail the inconclusive results for weight loss below. %
\begin{tablewide}
\begin{tab}{llrrcrr}
\multirow{2}{*}{Experiment} & \multirow{2}{*}{Keywords} & \multicolumn{2}{c}{\# ads (\# unique ads)} && \multicolumn{2}{c}{appearances}  \\
\cline{3-4} \cline{6-7}
 &  & removed & kept && removed & kept\\
\midrule
Dating          & dating, romance, relationship         & $952$ $(117)$ & $994$ $(123)$     && $34$ & $109$ \\
Weight Loss (1) & fitness                               & $1,461$ $(259)$ & $1,401$ $(240)$ && $21$ & $16$\\
Weight Loss (2) & fitness, health, fat, diet, exercise  & $1,803$ $(199)$ & $1,478$ $(192)$ && $2$ & $15$\\
\end{tab}
\caption{Setup for and ads from ad choice experiments.  All experiments used 10 blocks.  The same keywords are used to remove ad interests, as well as create the test statistic for permutation test.}
\label{tbl:ad-choice-ads}
\end{tablewide}
\newcommand{\antiarrayspace}{-1.5ex}
\begin{tablewide}
  \begin{tab}{lllllll} 
\multirow{2}{*}{Experiment} 
& \multirow{2}{*}{$\mbox{}\hspace{\antiarrayspace}\begin{array}{l}\text{Unadjusted}\\[\tightlinespace]\text{p-value}\end{array}$} 
& \multirow{2}{*}{$\mbox{}\hspace{\antiarrayspace}\begin{array}{l}\text{Bonferroni}\\[\tightlinespace]\text{p-value}\end{array}$} 
& \multirow{2}{*}{$\mbox{}\hspace{\antiarrayspace}\begin{array}{l}\text{Holm-Bonferroni}\\[\tightlinespace]\text{p-value}\end{array}$} 
& \multirow{2}{*}{$\mbox{}\hspace{\antiarrayspace}\begin{array}{l}\text{Unadjusted}\\[\tightlinespace]\text{flipped p-value}\end{array}$} 
& \multirow{2}{*}{$\mbox{}\hspace{\antiarrayspace}\begin{array}{l}\text{Bonferroni}\\[\tightlinespace]\text{flipped p-value}\end{array}$} 
& \multirow{2}{*}{$\mbox{}\hspace{\antiarrayspace}\begin{array}{l}\text{Holm-Bonferroni}\\[\tightlinespace]\text{flipped p-value}\end{array}$} \\
\\
\midrule
Dating          & $0.0076$ & $0.0152$ & $0.0456^*$ & $0.9970$  & $1.994$ & n/a\\
Weight Loss (2) & $0.18$   & $0.36$  & $0.9$     & $0.9371$  & $1.8742$ & n/a\\
Weight Loss (1) & $0.72$   & $1.44$  & n/a        & $0.3818$  & $0.7636$ & n/a\\
\end{tab}
\caption{P-values from ad choice experiments sorted by the (unflipped) p-value.  
The Bonferroni adjusted p-value is only adjusted for the two hypotheses tested within a single experiment (row).
The Holm-Bonferroni adjusts for all $6$ hypotheses.
$\mbox{}^*$ denotes statistically significant results under the Holm-Bonferroni method.}
\label{tbl:ad-choice-p-values}
\end{tablewide}

\paragraph{Weight Loss}
We induced an interest in weight loss by visiting \url{dietingsucks.blogspot.com}.
Afterwards, the agents in the experimental group removed the interests ``Fitness'' and ``Fitness Equipment and Accessories'', the only ones related to weight loss.
We then used a test statistic that counted the number of ads containing the keyword ``fitness''.
Interestingly, the test statistic was higher on the group with the interests removed, although not to a statistically significant degree.
We repeated the process with a longer keyword list and found that removing interests decreased test statistic this time, but also not to a statistically significant degree.

\section{Discussion and Conclusion}
\label{sec:disc}

Using AdFisher, we conducted 21 experiments using 17,370 %
agents that collected over 600,000 ads. 
Our experiments found instances of discrimination, opacity, and choice in targeted ads of Google.
Discrimination, is at some level, inherent to profiling: the point of profiling is to treat some people differently.  While customization can be helpful, we highlight a case where the customization appears inappropriate taking on the negative connotations of discrimination.  In particular, we found that males were shown ads encouraging the seeking of coaching services for high paying jobs more than females (\S\ref{sec:gender}).  

We do not, however, claim that any laws or policies were broken.
Indeed, Google's policies allow it to serve different ads based on gender.
Furthermore, we cannot determine whether Google, the advertiser, or complex interactions among them and others caused the discrimination (\S\ref{sec:scope}).
Even if we could, the discrimination might have resulted unintentionally from algorithms optimizing click-through rates or other metrics free of bigotry.
Given the pervasive structural nature of gender discrimination in society at large, blaming one party may ignore context and correlations that make avoiding such discrimination difficult.
More generally, we believe that no scientific study can demonstrate discrimination in the sense of \emph{unjust discrimination} since science cannot demonstrate normative statements (e.g.,~\cite{hume1738treatise})

Nevertheless, we are comfortable describing the results as ``discrimination''.  From a strictly scientific view point, we have shown discrimination in the non-normative sense of the word.  Personally, we also believe the results show discrimination in the normative sense of the word.
Male candidates getting more encouragement to seek coaching services for high-paying jobs could further the current gender pay gap (e.g.,~\cite{pew13pay}).
Thus, we do not see the found discrimination in our vision of a just society even if we are incapable of blaming any particular parties for this outcome.

Furthermore, we know of no justification for such customization of the ads in question.
Indeed, our concern about this outcome does not depend upon how the ads were selected.
Even if this decision was made solely for economic reasons, it would continue to be discrimination~\cite{zarsky14lawreview}.
In particular, we would remain concerned if the cause of the discrimination was an algorithm ran by Google and/or the advertiser automatically determining that males are more likely than females to click on the ads in question.
The amoral status of an algorithm does not negate its effects on society.

However, we also recognize the possibility that no party is at fault and such unjust effects may be inadvertent and difficult to prevent.
We encourage research developing tools that ad networks and advertisers can use to prevent such unacceptable outcomes (e.g.,~\cite{zemel13jmlr}).

Opacity occurs when a tool for providing transparency into how ads are selected and the profile kept on a person actually fails to provide such transparency.  Our experiment on substance abuse showed an extreme case in which the tool failed to show any profiling but the ad distributions were significantly different in response to behavior (\S\ref{sec:subs}).
In particular, our experiment achieved an adjusted p-value of $< 0.00005$, which is 1000 times more significant than the standard $0.05$ cutoff for statistical significance.
This experiment remained robust to variations showing a pattern of such opacity.

Ideally, tools, such as Ad Settings, would provide a complete representation of the profile kept on a person, or at least the portion of the profile that is used to select ads shown to the person.  Two people with identical profiles might continue to receive different ads due to other factors affecting the choice of ads such as A/B testing or the time of day.  However, systematic differences between ads shown at the same time and in the same context, such as those we found, would not exist for such pairs of people.

In our experiments testing transparency, we suspect that Google served the top ads as part of remarketing, but our blackbox experiments do not determine whether this is the case.
While such remarketing may appear less concerning than Google inferring a substance abuse issue about a person,
its highly targeted nature is worrisome particularly in settings with shared computers or shoulder surfing.
There is a need for a more inclusive transparency/control mechanism which encompasses remarketed ads as well.
Additionally, Google states that ``we prohibit advertisers from remarketing based on sensitive information, such as health information''~\cite{google-remarketing}. Although Google does not specify what they consider to be ``health information'', we view the ads as in violation of Google's policy, thereby raising the question of how Google should enforce its policies.

Lastly, we found that Google Ad Settings does provide the user with a degree of choice about the ads shown.  
In this aspect, the transparency/control tool operated as we expected.
Our tool, AdFisher, makes it easy to run additional experiments exploring the relations between Google's ads and settings. It can be extended to study other systems.  It's design ensures that it can run and analyze large scale experiments to find subtle differences.  
It automatically finds differences between large data sets produced by different groups of agents and explains the nature of those differences.
By completely automating the data analysis, we ensure that an appropriate statistical analysis determines whether these differences are statistically significant and sound conclusions.

AdFisher may have cost advertisers a small sum of money.
AdFisher never clicked on any ads to avoid per click fees, which can run over \$$4$~\cite{adgooroo}.
Its experiments may have caused per-impression fees, which run about \$$0.00069$~\cite{olejnik13ndss}.
In the billion dollar ad industry, its total effect was about \$$400$.

\section{Future Work}

We would like to extend AdFisher to study information flow on other advertising systems like Facebook, Bing, or Gmail. We would also like to analyze other kinds of ads like image or flash ads. We also plan to use the tool to detect price discrimination on sites like Amazon or Kayak, or find differences in suggested posts on blogs and news websites, based on past user behavior. We have already mentioned the interesting problem of how ad networks can ensure that their policies are respected by advertisers (\S\ref{sec:disc}).

We also like to assign blame where it is due.  However, doing so is often difficult.
For example, our view on blame varies based on why females were discriminated against in our gender and jobs experiment.
If Google allowed the advertiser to easily discriminate, we would blame both.
If the advertiser circumvented Google's efforts to prevent such discrimination by targeting correlates of gender, we would blame just the advertiser.
If Google decided to target just males with the ad on its own, we would blame just Google.
While we lack the access needed to make this determination, both Google and the advertiser have enough information to audit the other with our tool.

As another example, consider the results of opacity after visiting substance abuse websites.
While we suspect, remarketing is the cause, it is also possible that Google is targeting users without the rehab center's knowledge.  
In this case, it would remain unclear as to whether Google is targeting users as substance abusers or due to some other content correlated with the webpages we visited to simulate an interest in substance abuse.
We would like to find ways of controlling for these confounding factors.

For these reasons, we cannot claim that Google has violated its policies.  
In fact, we consider it more likely that Google has lost control over its massive, automated advertising system. 
Even without advertisers placing inappropriate bids, large-scale machine learning can behave in unexpected ways.  
With this in mind, we hope future research will examine how to produce machine learning algorithms that automatically avoid discriminating against users in unacceptable ways and automatically provide transparency to users. %

\paragraph{Acknowledgements}
We thank Jeannette M.\ Wing for helpful discussions about this work.
We thank Augustin Chaintreau, Roxana Geambasu, Qiang Ma, Latanya Sweeney, and Craig E.\ Wills for providing additional information about their works.
We thank the reviewers of this paper for their helpful comments.
This research was supported by the National Science Foundation (NSF) grants CCF0424422 and CNS1064688.
The views and conclusions contained in this document are those of the authors and should not be interpreted as representing the official policies, either expressed or implied, of any sponsoring institution, the U.S.\ government or any other entity.

\bibliographystyle{acm}
\bibliography{blackbox}

\begin{thebibliography}{10}

\bibitem{abdi07enc}
{\sc Abdi, H.}
\newblock Bonferroni and \v{S}id\'ak corrections for multiple comparisons.
\newblock In {\em Encyclopedia of Measurement and Statistics}, N.~J. Salkind,
  Ed. Sage, 2007.

\bibitem{adgooroo}
{\sc Adgooroo}.
\newblock Adwords cost per click rises 26\% between 2012 and 2014.
\newblock
  \url{http://www.adgooroo.com/resources/blog/adwords-cost-per-click-rises-26-between-2012-and-2014/}.
\newblock Accessed Nov.~21, 2014.

\bibitem{alexa}
{\sc Alexa}.
\newblock Is popularity in the top sites by category directory based on traffic
  rank?
\newblock \url{https://support.alexa.com/hc/en-us/articles/200461970}.
\newblock Accessed Nov.~21, 2014.

\bibitem{balebako12w2sp}
{\sc Balebako, R., Leon, P., Shay, R., Ur, B., Wang, Y., and Cranor, L.}
\newblock Measuring the effectiveness of privacy tools for limiting behavioral
  advertising.
\newblock In {\em Web 2.0 Security and Privacy Workshop\/} (2012).

\bibitem{barford14www}
{\sc Barford, P., Canadi, I., Krushevskaja, D., Ma, Q., and Muthukrishnan, S.}
\newblock Adscape: Harvesting and analyzing online display ads.
\newblock In {\em Proceedings of the 23rd International Conference on World
  Wide Web\/} (2014), International World Wide Web Conferences Steering
  Committee, pp.~597--608.

\bibitem{bishop06pattern}
{\sc Bishop, C.~M.}
\newblock {\em Pattern Recognition and Machine Learning}.
\newblock Springer, 2006.

\bibitem{clopper34biometrika}
{\sc Clopper, C.~J., and Pearson, E.~S.}
\newblock The use of confidence or fiducial limits illustrated in the case of
  the binomial.
\newblock {\em Biometrika 26}, 4 (1934), 404--413.

\bibitem{datta14arxiv}
{\sc Datta, A., Tschantz, M.~C., and Datta, A.}
\newblock Automated experiments on ad privacy settings: A tale of opacity,
  choice, and discrimination.
\newblock Tech. Rep. arXiv:1408.6491v1, ArXiv, 2014.

\bibitem{datta15pets}
{\sc Datta, A., Tschantz, M.~C., and Datta, A.}
\newblock Automated experiments on ad privacy settings: A tale of opacity,
  choice, and discrimination.
\newblock In {\em Proceedings on Privacy Enhancing Technologies (PoPETs)\/}
  (2015), De Gruyter Open.

\bibitem{englehardt14man}
{\sc Englehardt, S., Eubank, C., Zimmerman, P., Reisman, D., and Narayanan, A.}
\newblock Web privacy measurement: Scientific principles, engineering platform,
  and new results.
\newblock Manuscript posted at
  \url{http://randomwalker.info/publications/WebPrivacyMeasurement.pdf}, 2014.
\newblock Accessed Nov. 22, 2014.

\bibitem{bigdata14whitehouse}
{\sc {Executive Office of the President}}.
\newblock Big data: Seizing opportunities, preserving values.
\newblock Posted at
  \url{http://www.whitehouse.gov/sites/default/files/docs/big_data_privacy_report_may_1_2014.pdf},
  2014.
\newblock Accessed Jan. 26, 2014.

\bibitem{fisher35doe}
{\sc Fisher, R.~A.}
\newblock {\em The Design of Experiments}.
\newblock Oliver \& Boyd, 1935.

\bibitem{good05book}
{\sc Good, P.}
\newblock {\em Permutation, Parametric and Bootstrap Tests of Hypotheses}.
\newblock Springer, 2005.

\bibitem{google-ad-settings-help}
{\sc Google}.
\newblock About ads settings.
\newblock \url{https://support.google.com/ads/answer/2662856}.
\newblock Accessed Nov.~21, 2014.

\bibitem{google-remarketing}
{\sc Google}.
\newblock Google privacy and terms.
\newblock \url{http://www.google.com/policies/technologies/ads/}.
\newblock Accessed Nov.~22, 2014.

\bibitem{google-privacy}
{\sc Google}.
\newblock Privacy policy.
\newblock \url{https://www.google.com/intl/en/policies/privacy/}.
\newblock Accessed Nov.~21, 2014.

\bibitem{greenland86epidemiology}
{\sc Greenland, S., and Robins, J.~M.}
\newblock Identifiability, exchangeability, and epidemiological confounding.
\newblock {\em International Journal of Epidemiology 15}, 3 (1986), 413--419.

\bibitem{guha10imc}
{\sc Guha, S., Cheng, B., and Francis, P.}
\newblock Challenges in measuring online advertising systems.
\newblock In {\em Proceedings of the 10th ACM SIGCOMM Conference on Internet
  Measurement\/} (2010), pp.~81--87.

\bibitem{holm79scandinavian}
{\sc Holm, S.}
\newblock A simple sequentially rejective multiple test procedure.
\newblock {\em Scandinavian Journal of Statistics 6}, 2 (1979), 65--70.

\bibitem{hume1738treatise}
{\sc Hume, D.}
\newblock {\em A Treatise of Human Nature: Being an Attempt to Introduce the
  Experimental Method of Reasoning into Moral Subjects}.
\newblock 1738.
\newblock Book III, part I, section I.

\bibitem{jensen92phd}
{\sc Jensen, D.~D.}
\newblock {\em Induction with Randomization Testing: Decision-oriented Analysis
  of Large Data Sets}.
\newblock PhD thesis, Sever Institute of Washington University, 1992.

\bibitem{scipy}
{\sc Jones, E., Oliphant, T., Peterson, P., et~al.}
\newblock {SciPy}: Open source scientific tools for {Python}, 2001.
\newblock \url{http://www.scipy.org/}.

\bibitem{lecuyer14usenix}
{\sc L\'ecuyer, M., Ducoffe, G., Lan, F., Papancea, A., Petsios, T., Spahn, R.,
  Chaintreau, A., and Geambasu, R.}
\newblock {XR}ay: Increasing the web's transparency with differential
  correlation.
\newblock In {\em Proceedings of the USENIX Security Symposium\/} (2014).

\bibitem{liu13hotnets}
{\sc Liu, B., Sheth, A., Weinsberg, U., Chandrashekar, J., and Govindan, R.}
\newblock {AdReveal}: Improving transparency into online targeted advertising.
\newblock In {\em Proceedings of the Twelfth ACM Workshop on Hot Topics in
  Networks\/} (2013), ACM, pp.~12:1--12:7.

\bibitem{mayer12sp}
{\sc Mayer, J.~R., and Mitchell, J.~C.}
\newblock Third-party web tracking: Policy and technology.
\newblock In {\em IEEE Symposium on Security and Privacy\/} (2012),
  pp.~413--427.

\bibitem{microsoft-choice}
{\sc Microsoft}.
\newblock Microsoft personalized ad preferences.
\newblock \url{http://choice.microsoft.com/en-us/opt-out}.
\newblock Accessed Nov.~21, 2014.

\bibitem{mitchell97book}
{\sc Mitchell, T.~M.}
\newblock {\em Machine Learning}.
\newblock McGraw-Hill, 1997.

\bibitem{olejnik13ndss}
{\sc Olejnik, L., Minh-Dung, T., and Castelluccia, C.}
\newblock Selling off privacy at auction.
\newblock In {\em Network and Distributed System Security Symposium (NDSS)\/}
  (2013), The Internet Society.

\bibitem{scikit-learn}
{\sc Pedregosa, F., Varoquaux, G., Gramfort, A., Michel, V., Thirion, B.,
  Grisel, O., Blondel, M., Prettenhofer, P., Weiss, R., Dubourg, V.,
  Vanderplas, J., Passos, A., Cournapeau, D., Brucher, M., Perrot, M., and
  Duchesnay, E.}
\newblock Scikit-learn: Machine learning in {P}ython.
\newblock {\em Journal of Machine Learning Research 12\/} (2011), 2825--2830.

\bibitem{pew13pay}
{\sc {Pew Research Center's Social and Demographic Trends Project}}.
\newblock On pay gap, millennial women near parity --- for now: Despite gains,
  many see roadblocks ahead, 2013.

\bibitem{sweeney13cacm}
{\sc Sweeney, L.}
\newblock Discrimination in online ad delivery.
\newblock {\em Commun. ACM 56}, 5 (2013), 44--54.

\bibitem{tschantz14arxiv}
{\sc Tschantz, M.~C., Datta, A., Datta, A., and Wing, J.~M.}
\newblock A methodology for information flow experiments.
\newblock Tech. Rep. arXiv:1405.2376v1, ArXiv, 2014.

\bibitem{ur12soups}
{\sc Ur, B., Leon, P.~G., Cranor, L.~F., Shay, R., and Wang, Y.}
\newblock Smart, useful, scary, creepy: Perceptions of online behavioral
  advertising.
\newblock In {\em Proceedings of the Eighth Symposium on Usable Privacy and
  Security\/} (2012), ACM, pp.~4:1--4:15.

\bibitem{wills12wpes}
{\sc Wills, C.~E., and Tatar, C.}
\newblock Understanding what they do with what they know.
\newblock In {\em Proceedings of the 2012 ACM Workshop on Privacy in the
  Electronic Society\/} (2012), pp.~13--18.

\bibitem{yahoo-help}
{\sc Yahoo!}
\newblock Ad interest manager.
\newblock
  \url{https://info.yahoo.com/privacy/us/yahoo/opt_out/targeting/details.html}.
\newblock Accessed Nov.~21, 2014.

\bibitem{zarsky14lawreview}
{\sc Zarsky, T.~Z.}
\newblock Understanding discrimination in the scored society.
\newblock {\em Washington Law Review 89\/} (2014), 1375--1412.

\bibitem{zemel13icml}
{\sc Zemel, R., Wu, Y., Swersky, K., Pitassi, T., and Dwork, C.}
\newblock Learning fair representations.
\newblock In {\em Proceedings of the 30th International Conference on Machine
  Learning (ICML-13)\/} (May 2013), S.~Dasgupta and D.~Mcallester, Eds.,
  vol.~28, JMLR Workshop and Conference Proceedings, pp.~325--333.

\bibitem{zemel13jmlr}
{\sc Zemel, R.~S., Wu, Y., Swersky, K., Pitassi, T., and Dwork, C.}
\newblock Learning fair representations.
\newblock In {\em Proceedings of the 30th International Conference on Machine
  Learning\/} (2013), vol.~28 of {\em JMLR: W\&CP}, JMLR.org, pp.~325--333.

\end{thebibliography}

\appendix

\section{Details of Methodology}\label{app:stat}

Let the units be arranged in a vector $\vec{u}$ of length $n$.  Let $\vec{t}$ be a \emph{treatment vector}, a vector of length $n$ whose entries are the treatments that the experimenter wants to apply to the units.  In the case of just two treatments, $\vec{t}$ can be half full of the first treatment and half full of the second.  Let $a$ be an \emph{assignment} of units to treatments, a bijection that maps each entry of $\vec{u}$ to an entry in $\vec{t}$.  That is, an assignment is a permutation on the set of indices of $\vec{u}$ and $\vec{t}$.

The result of the experiment is a vector of observations $\vec{y}$ where the $i$th entry of $\vec{y}$ is the response measured for the unit assigned to the $i$th treatment in $\vec{t}$ by the assignment used.
In a randomized experiment, such as those AdFisher runs, the actual assignment used is selected at random uniformly over some set of possible assignments $\mc{A}$.

Let $s$ be a test statistic of the observations of the units.  That is $s : \mc{Y}^n \to \mc{R}$ where $\mc{Y}$ is the set of possible observations made over units, $n$ is the number of units, and $\mc{R}$ is the range of $s$.  We require $\mc{R}$ to be ordered numbers such as the natural or real numbers.
We allow $s$ to treat its arguments differently, that is, the order in which the observations are passed to $s$ matters.

If the null hypothesis is true, then we would expect the value of $s$ to be the same under every permutation of the arguments %
since the assignment of units to treatments should not matter under the null hypothesis.
This reasoning motivates the permutation test.
The value produced by a (one-tailed signed) permutation test given observed responses $\vec{y}$ and a test statistic $s$ is
\begin{align}
\frac{|\set{a \in \mc{A}}{s(\vec{y}) \leq s(a(\vec{y}))}|}{|\mc{A}|}
= \frac{1}{|\mc{A}|} \sum_{a \in \mc{A}} I[s(\vec{y}) \leq s(a(\vec{y}))] \label{eqn:pt}
\end{align}
where the assignments in $\mc{A}$ only swaps nearly identical units and $I[\cdot]$ returns $1$ if its argument is true and $0$ otherwise.

\paragraph{Blocking}
For the blocking design, the set of units $\mc{U}$ is partitioned into $k$ blocks $\mc{B}_1$ to $\mc{B}_k$.  In our case, all the blocks have the same size.  Let $|\mc{B}_i| = m$ for all $i$.
The set of assignments $\mc{A}$ is equal to the set of functions from $\mc{U}$ to $\mc{U}$ that are permutations not mixing up blocks.  That is, $a$ such that for all $i$ and all $u$ in $\mc{B}_i$, $a(u) \in \mc{B}_i$.  Thus, we may treat $\mc{A}$ as $k$ permutations, one for each $\mc{B}_i$.  Thus, $\mc{A}$ is isomorphic to $\bigtimes_{i = 1}^{k} \Pi(\mc{B}_i)$ where
$\Pi(\mc{B}_i)$ is the set of all permutations over $\mc{B}_i$.  Thus, $|\bigtimes_{i = 1}^{k} \Pi(\mc{B}_i)| = (m!)^k$.
Thus, \eqref{eqn:pt} can be computed as
\begin{align}
\frac{1}{(m!)^k} \sum_{a \in \bigtimes_{i = 1}^{k} \Pi(\mc{B}_i)} I[s(\vec{y}) \leq s(a(\vec{y}))] 
\label{eqn:pt-block}
\end{align}

\paragraph{Sampling}
Computing \eqref{eqn:pt-block} can be difficult when the set of considered arrangements is large.  One solution is to randomly sample from the assignments $\mc{A}$.  Let $\mc{A}'$ be a random subset of $\mc{A}$.  We then use the approximation
\begin{align}
\frac{1}{|\mc{A}'|} \sum_{a \in \mc{A}'} I[s(\vec{y}) \leq s(a(\vec{y}))] \label{eqn:pt-approx}
\end{align}

\paragraph{Confidence Intervals}
Let $\hat{P}$ be this approximation and $p$ be the true value of \eqref{eqn:pt-block}.  $p$ can be understood as the frequency of arrangements that yield large values of the test statistic where \emph{largeness} is determined to be at least as large as the observed value $s(\vec{y})$.  That is, the probability that a randomly selected arrangement will yield a large value is $p$.  $\hat{P}$ is the frequency of seeing large values in the $|\mc{A}'|$ sampled arrangements.  Since the arrangements in the sample were drawn uniformly at random from $\mc{A}$ and each draw has probability $p$ of being large, the number of large values will obey the binomial distribution.  Let us denote this value as $L$.
and $|\mc{A}'|$ as $n$.
Since $\hat{P} = L/n$, $\hat{p}*n$ also obeys the binomial distribution.  Thus,
\begin{align}
\Pr[ \hat{P} = \hat{p} \given n, p] = \binom{n}{\hat{p}n} p^{\hat{p}n} (1-p)^{(1-\hat{p})n}
\end{align}

Thus, we may use a binomial proportion confidence interval.  
We use the Clopper-Pearson interval~\cite{clopper34biometrika}.

\paragraph{Test Statistic}
The statistic we use is based on a classifier $c$.  
Let $c(y_i) = 1$ mean that $c$ classifiers the $i$th observation as having come from the experimental group and $c(y_i) = 0$ as from the control group.  Let $\neg(0) = 1$ and $\neg(1) = 0$.  Let $\vec{y}$ be ordered so that all of the experimental group comes first.
The statistic we use is
\begin{align}
s(\vec{y}) &= \sum_{i = 1}^{n/2} c(y_i) + \sum_{i = n/2 + 1}^{n} \neg c(y_i)
\end{align}
This is the number correctly classified.

\section{Holm-Bonferroni Correction}
\label{app:correction}

The Holm-Bonferroni Correction starts by ordering the hypotheses in a family from the hypothesis with the smallest (most significant) p-value $p_1$ to the hypothesis with the largest (least significant) p-value $p_m$~\cite{holm79scandinavian}.
For a hypothesis $H_k$, its unadjusted p-value $p_k$ is compared to an adjusted level of significance $\alpha'_k = \frac{\alpha}{m + 1 - k}$ 
where $\alpha$ is the unadjusted level of significance ($0.05$ in our case), $m$ is the total number of hypotheses in the family, and $k$ is the index of hypothesis in the ordered list (counting from $1$ to $m$).
Let $k^\dagger$ be the lowest index $k$ such that $p_k > \alpha'_k$.
The hypotheses $H_k$ where $k < k^\dagger$ are accepted as having statistically significance evidence in favor of them (more technically, the corresponding null hypotheses are rejected).
The hypotheses $H_k$ where $k \geq k^\dagger$ are not accepted as having significant evidence in favor of them (their null hypotheses are not rejected).

We report adjusted p-values to give an intuition about the strength of evidence for a hypothesis.
We let $p'_k = p (m + 1 - k)$ be the adjusted p-value for $H_k$ provided $k < k^\dagger$ since $p_k > \alpha'_k$ iff $p'_k > \alpha$.
Note that the adjusted p-value depends not just upon its unadjusted value but also upon its position in the list.
For the remaining hypotheses, we provide no adjusted p-value since their p-values are irrelevant to the correction beyond how they order the list of hypotheses.

\end{document}